\documentclass[twocolumn,showpacs,amsmath,amssymb,prb,floatfix,aps]{revtex4-1}
\usepackage{graphicx}
\usepackage{dcolumn}
\usepackage{bm}
\begin{document}
\title{Periodic Landau gauge and 
Quantum Hall effect in twisted bilayer graphene 
}
\author{Yasumasa Hasegawa$^1$
and Mahito Kohmoto$^2$}
\affiliation
{$^1$Department of Material Science, Graduate School of Material Science, 
University of Hyogo, \\
3-2-1 Kouto, Kamigori, Hyogo, 678-1297, Japan \\
$^2$Institute for Solid State Physics, University of Tokyo, 
5-1-5 Kashiwanoha, Kashiwa, Chiba 277-8581, Japan
}
%
\date{April 26, 2013, \today}

\begin{abstract}
Energy versus magnetic field
(Hofstadter butterfly diagram) 
in  twisted bilayer graphene is studied theoretically.
If we take the usual Landau gauge, we cannot take a finite 
periodicity even when the magnetic flux through a supercell
is a rational number.
We show that the \textit{periodic} Landau gauge, 
which has the periodicity in 
one direction, makes it possible to obtain 
the Hofstadter butterfly diagram. 
Since a supercell can be large,
magnetic flux through a supercell normalized by the
flux quantum can be a fractional number with a small denominator,
even when a magnetic field is not extremely strong.
As a result, quantized Hall conductance can
be a solution of 
the Diophantine equation which cannot be obtained 
by the approximation of the linearized energy 
dispersion near the Dirac points.
\end{abstract}

\maketitle

\section{Introduction}
\begin{figure}[b]
%
\begin{center}
\vspace*{0.4cm} 
\includegraphics[width=0.41\textwidth]{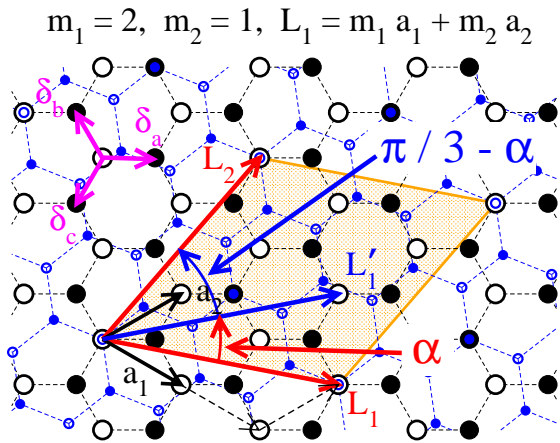} 
\end{center}
\caption{(color online).
Twisted bilayer graphene with $(m_1,m_2)=(2,1)$.
Large (small) open and filled circles are A and B sublattice in 
the first (second) 
layer, respectively. The second layer is rotated by $\pi/3 -\alpha$.
The orange area is the supercell.
}
\label{figfig21}
\end{figure}
Two dimensional electron systems are
realized in graphene.\cite{Novoselov2004}
Bilayer graphene\cite{McCann2006PRL} and twisted 
bilayer graphene\cite{Shallcross2010} have been shown 
to have many interesting properties and been studied extensively.
The quantum Hall effects in single layer 
graphene\cite{Novoselov2005,Zhang2005}
and bilayer graphene are well understood
by the energy versus magnetic field,
which is known as the Hofstadter butterfly diagram,
 in the tight binding models
of honeycomb lattice\cite{Hasegawa2006,Hatsugai2006,Dietl2008},
and on Bernal-stacked bilayer graphene\cite{Nemec2007}. 
The quantized value of the Hall
conductance is obtained by the solution of the Diophantine 
equation.\cite{Thouless1982,Kohmoto1985,Kohmoto1989}
Near half filled case, 
the quantum Hall conductance 
is given by a solution of the Diophantine equation Eq.~(\ref{eqDiophantine})
with $s_r=1$ ($s_r=2$) in single layer (bilayer) graphene.

Twisted bilayer graphene attracts much attention 
recently. When two layers are twisted in a commensurate way,
a supercell becomes 
large with a moir\`e pattern 
\cite{Lopes2007,Hass2008,Shallcross2008,Shallcross2010,Mele2010} 
(see Fig.~\ref{figfig21})
and the velocity at the Dirac points is shown to become smaller when
the rotation angle $\alpha$ is small.\cite{Trambly2010,BistritzerPNAS2011}

Quantum Hall effect and the Hofstadter butterfly diagram
in moir\`e superlattices have 
been observed experimentally in  single layer graphene 
on hexagonal boron nitride (hBN)\cite{Ponomarenko2012}
and Bernal-stacked bilayer graphene on hBN.\cite{Dean2012}

Electronic structure in twisted bilayer graphene 
in a uniform magnetic field 
has been studied
only by taking the linearized energy dispersion near the Dirac 
points\cite{Lee2011,BistritzerPRB2011,Moon2012} or by the Lanczos algorithm 
applied to large systems in real space\cite{Wang2012}.
The whole lattice structure has not been taken into account
when we take the linearized energy dispersion. 
Exact band structure is difficult to obtain by the Lanczos
algorithm due to the finite-size effects and  numerical errors.

When we use the usual Landau gauge
in twisted bilayer graphene with long range
hoppings, there are no periodicity in the phase factor of  hoppings. 
In that case 
we cannot obtain the Hofstadter butterfly diagram.
In this paper we show that we can recover the
periodicity when a magnetic flux through a supercell is
a rational number, if we use the periodic Landau
 gauge.
As far as we know, a special choice of gauge was first used 
to study the system
of the $4 \times 4$ square lattice with periodic boundary 
conditions in the presence
of the uniform magnetic flux $p/16$ with 
$p=1, 2, 3, \cdots$\cite{Poilblanc1990}
(if a usual Landau gauge is used in that system,
only magnetic flux $p/4$ is allowed).
 String gauge, which is obtained by adding the flux line
with a flux quantum, has
been introduced to study the periodic system 
in a magnetic field.\cite{Hatsugai1999} 
The periodic vector potential (equivalent to the periodic Landau gauge)
has been introduced to study the Schr\"odinger equation with 
 periodic potential in a uniform magnetic field\cite{Trellakis2003},
and it has been applied to study the tight binding model
in Bernal stacked bilayer graphene\cite{Nemec2007}. 
Another approach by using Fourier transform has been proposed
for the periodic system in a magnetic field.\cite{Cai2004}
However, the periodic Landau gauge has not been used to study
 twisted bilayer graphene in 
a magnetic field.
By virtue of the periodic Landau gauge
we can calculate the energy spectrum in twisted 
 bilayer graphene 
in a magnetic field 
in a similar way as
in single layer graphene\cite{Hasegawa2006,Hatsugai2006,Dietl2008}
 or Bernal stacked bilayer graphene\cite{Nemec2007}.
We obtain very rich Hofstadter diagrams, 
which have not been obtained 
in previous studies\cite{Lee2011,BistritzerPRB2011,Moon2012,Wang2012}.
We find many energy gaps near half filled case, 
which are indexed by integers
given by a solution of the Diophantine equation as
 $(s_r, t_r) = (2 n_0, 2)$, $(2 n_0 \pm 2, 0)$, 
 $(2n_0 \pm 2, 2)$,  $(2n_0 \pm 4, 2)$, $(2n_0 \pm 6, 2)$  $\cdots$,
where $n_0$ is the number of the A site in the first layer
in the supercell.
 The quantized value
of the Hall conductance is obtained by $t_r$.

In Section II we define the twisted bilayer graphene with
commensurate twisted angle. In Section III the tight binding model
and the periodic Landau gauge are explained. 
In Section IV we show the Hofstadter butterfly diagram  
and study the quantized Hall conductances, 
which are obtained by the Diophantine equation. 
We give the summary in Section V.
The detailed explanation of the periodic Landau gauge 
in the square lattice is given in Appendix A. 
The periodic Landau gauge in the twisted bilayer graphene
is discussed in Appendix B.
\section{twisted bilayer graphene}
In a unit cell in each layer there are two
sites, A and B, which form triangular lattices respectively. 
We define unit vectors as
\begin{equation}
 \mathbf{a}_1 = a \left( \begin{array}{c}
\frac{\sqrt{3}}{2} \\ -\frac{1}{2} \end{array} \right), 
\end{equation}
and
\begin{equation}
 \mathbf{a}_2 = R_{\pi/3} \mathbf{a}_1 =a \left( \begin{array}{c}
\frac{\sqrt{3}}{2} \\ \frac{1}{2} \end{array} \right),
\end{equation}
where $a$ is the lattice constant and $R_{\pi/3}\mathbf{a}_1$  is the 
$\pi/3$ rotated vector of $\mathbf{a}_1$.
Hereafter we take $a=1$ for simplicity. 
The reciprocal lattice vectors are given by
\begin{align}
 \mathbf{G}_1 &= 2 \pi \left( \begin{array}{c}
\frac{1}{\sqrt{3}} \\ -1 \end{array} \right), \\
 \mathbf{G}_2 &= 2 \pi \left( \begin{array}{c}
\frac{1}{\sqrt{3}} \\ 1 \end{array} \right).
\end{align}
In the first layer, sites in the A sublattice are given by
sets of two integers $(j_1, j_2)$ as
\begin{equation}
\mathbf{r}^{A}_{j_1,j_2} = j_1 \mathbf{a}_1 + j_2 \mathbf{a}_2.
\end{equation}
%
Three vectors connecting nearest neighbor sites 
in the first layer are
\begin{align}
 \boldsymbol{\delta}_a &= \left( \begin{array}{c}
    \frac{\sqrt{3}}{3} \\ 0            \end{array} \right).\\
 \boldsymbol{\delta}_b &= \left( \begin{array}{c}
   -\frac{\sqrt{3}}{6} \\ \frac{1}{2} \end{array} \right), \\
 \boldsymbol{\delta}_c &= \left( \begin{array}{c}
   -\frac{\sqrt{3}}{6} \\  -\frac{1}{2} \end{array} \right).
\end{align}
Sites in the B sublattice in the first layer are given by
\begin{equation}
\mathbf{r}^{B}_{j_1,j_2} = \mathbf{r}^{A}_{j_1,j_2} + \boldsymbol{\delta}_a.
\end{equation}
The AB (Bernal) stacking of bilayer graphene is obtained by rotating
the second layer around  one of the A site in the first layer
by the angle $(2n+1) \pi/3$, where $n$ is an integer.
In this case the A sublattice in the second layer is just above the 
A sublattice in the first layer, but the B sublattice in the
second layer is on the center of the hexagon in the first layer.
The same stacking is obtained by translating the second layer 
by $-\boldsymbol{\delta}_a$,
$-\boldsymbol{\delta}_b$ or $-\boldsymbol{\delta}_c$.
When the rotation angle is  $2n \pi/3$, we obtain the AA stacking, i.e., 
all sites in the second layer are on the sites in the first layer.
We obtain twisted bilayer graphene, when the rotation angle is neither
$(2n+1) \pi/3$ nor $2n \pi/3$.

When twisted bilayer graphene has supercell with finite number of sites,
it is called commensurate twisted bilayer graphene.
We construct commensurate twisted bilayer graphene as follows;
Since there is six-fold symmetry in twisted bilayer graphene, 
we can take a supercell as a diamond with the angle $\pi/3$ 
as shown in Fig.~\ref{figfig21}.
We define unit vectors of superlattice with two integers $m_1$ and $m_2$
($m_1 \neq 0$, $m_2 \neq 0$, and $|m_1| \neq |m_2|$):
\begin{align}
 \mathbf{L}_1 &= \mathbf{r}^{A}_{m_1,m_2}
 =m_1 \mathbf{a}_1 + m_2 \mathbf{a}_2,\\
 \mathbf{L}_2 &= R_{\pi/3}\mathbf{L}_1 
=\mathbf{r}^{A}_{-m_2, (m_1+m_2)}.
\end{align}
Twisted bilayer graphene with 
$(m_1,m_2)=(2,1)$ is shown in Fig.~\ref{figfig21}.
Since
\begin{equation}
  \mathbf{a}_1 \cdot \mathbf{a}_2 = \frac{1}{2},
\end{equation}
we obtain 
\begin{equation}
 |\mathbf{L}_1|^2=|\mathbf{L}_2|^2
 = m_1^2+m_2^2+m_1 m_2 \equiv n_0.
\end{equation}
The area of a supercell is given by
\begin{equation}
 S = |\mathbf{L}_1| |\mathbf{L}_2| \sin \frac{\pi}{3} 
 = \frac{\sqrt{3}}{2} n_0.
\end{equation}
There is another site in the supercell that has the same 
distance from the origin as $|\mathbf{L_1}|$, which 
we define $\mathbf{L}_1'$ as shown in Fig.~\ref{figfig21}:
\begin{equation}
  \mathbf{L}_1' = \mathbf{r}_{m_2,m_1}^A 
 = m_2 \mathbf{a}_1+m_1 \mathbf{a}_2.
\end{equation}
We define $\alpha$ by 
the angle between the vectors $\mathbf{L}_1'$ and $\mathbf{L}_1$.
Since
\begin{equation}
  \mathbf{L}_1' \cdot \mathbf{L}_1 = \frac{1}{2} 
(m_1^2 + 4 m_1 m_2 + m_2^2),
\end{equation}
we obtain
\begin{equation}
  \cos \alpha = \frac{\mathbf{L}_1' \cdot \mathbf{L}_1}%
{|\mathbf{L}_1'| |\mathbf{L}_1|}=
\frac{m_1^2+4 m_1 m_2 + m_2^2}{2 (m_1^2 + m_1 m_2 + m_2^2)}.
\label{eqalpha}
\end{equation}
Then we obtain twisted bilayer graphene by rotating
the second layer with the angle $\pi/3-\alpha$ to move the
vector $\mathbf{L}_1'$ into $\mathbf{L}_2$. 
We obtain another type of twisted bilayer graphene when we 
rotate the second layer
by the angle $-\alpha$. In this paper we take the rotation 
angle $\pi/3-\alpha$ to obtain the Bernal stacking when $\alpha \to 0$.

The Bravais lattice of twisted bilayer graphene is 
the $-\alpha/2$-tilted two-dimensional triangular lattice 
with the primitive vectors
$\mathbf{L}_1$ and $\mathbf{L}_2$.
A supercell has
$n_0$ sites in A and B sublattice in each layer, hence $4n_0$ sites.

\section{tight binding model and periodic Landau gauge}
We consider tight binding models in a uniform magnetic field.
Spin is not taken into account. 
When a magnetic field is applied, 
the hopping $t_{ij}$ between sites $\mathbf{r}_i$
and $\mathbf{r}_j$
($\mathbf{r}_i$ and $\mathbf{r}_j$ are on the same layer
or on different layers)
has the factor $\exp(i \theta_{ij})$ with
a phase $\theta_{ij}$ given by
\begin{equation}
 \theta_{ij} = \frac{2\pi}{\phi_0}
\int_{\mathbf{r}_i}^{\mathbf{r}_j} \mathbf{A}
\cdot  d\boldsymbol{\ell},
\label{eqtheta}
\end{equation}
where
$\mathbf{A}$ is a vector potential and
\begin{equation}
 \phi_0=\frac{ch}{e},
\end{equation}
is the flux quantum with charge $e$, the speed of light $c$ and 
the Planck constant $h$.
The Hamiltonian is 
\begin{align}
 \mathcal{H} 
  &= -\sum_{(i,j)}(e^{i \theta_{ij}}t_{ij} 
  c^{\dagger}_{j} c_{i} + h.c.),
\end{align}
where $c^{\dagger}_{i}$ and $c_{i}$ are the creation 
and annihilation operators at site $i$, respectively.
We take the approximation\cite{Nakanishi2001,Trambly2010,Moon2012},
\begin{equation}
  t_{ij}= \left\{
\begin{array}{l}
t \exp\left(- \cfrac{d-|\boldsymbol{\delta_a}|}{\delta} \right)\\
  \hspace{1.2cm}  \mbox{ if sites $i$ and $j$ are on the same layer}\\
t_{12} \exp \left(-\cfrac{d-d_0}{\delta_{12}} \right) \\
  \hspace{1.2cm}  \mbox{ if sites $i$ and $j$ are on different layers}
\end{array} \right. ,
\end{equation}
where $d=|\mathbf{r}_i-\mathbf{r}_j|$ 
is the distance between sites $i$ and $j$
and $d_0$ is the distance between layers.
When we take $\delta \to 0$ and $t_{12}=0$,
we obtain two independent layers of honeycomb lattice 
with only nearest-neighbor hoppings.
When $t_{12}$, $\delta$ and $\delta_{12}$ are finite,
we obtain twisted bilayer graphene with finite range hoppings.
Interlayer hoppings are 
not restricted to the perpendicular direction.

The energy is independent of the sign of the interlayer hoppings $t_{12}$,
since we obtain the same Hamiltonian by changing the sing of $t_{12}$,
and the signs of $c_i^{\dagger}$ and $c_i$ in 
the second layer simultaneously.

Even if the flux per supercell is an integer times the flux quantum
$\phi_0$, the phase factor $\theta_{ij}$ is not periodic
with modulus $2\pi$, if we use the usual Landau gauge
($\mathbf{A}=Hx \hat{\mathbf{y}}$).
For single layer graphene with only nearest-neighbor hoppings, 
we could take a special gauge\cite{Hasegawa2006,Hatsugai2006,Rhim2012},
in which the phase factor appears only in the links for one of the
three directions, $\boldsymbol{\delta}_a$,
$\boldsymbol{\delta}_b$ or $\boldsymbol{\delta}_c$.
However, such choice of gauge is not possible 
for twisted bilayer graphene.

In this paper we study the energy spectrum in the
twisted bilayer graphene in magnetic field by using the 
periodic Landau gauge, which is essentially the same as
the gauge used by Nemec and Cuniberti\cite{Nemec2007}
 to study the Bernal stacked
bilayer graphene.
We explain the periodic Landau gauge for the square lattice 
in Appendix~\ref{labAppA}.
The generalization to the non-square lattice is given in  
Appendix~\ref{labAppB}.

When flux through a supercell is 
\begin{equation}
 \Phi= SH = \frac{p}{q} \phi_0,
\end{equation}
where $p$ and $q$ are mutually prime integers
in twisted bilayer
graphene with commensurate twisted angle (Eq.(\ref{eqalpha})),
 energy spectrum 
is obtained by the eigenvalues of $(4 n_0 q) \times (4 n_0 q)$ matrix,
as in the case of the single layer graphene where it is obtained
by  the eigenvalues of $(2 q) \times (2 q)$ matrix.\cite{Hasegawa2006} 

\begin{figure}[bt]
%
\begin{center}
\includegraphics[width=0.45\textwidth]{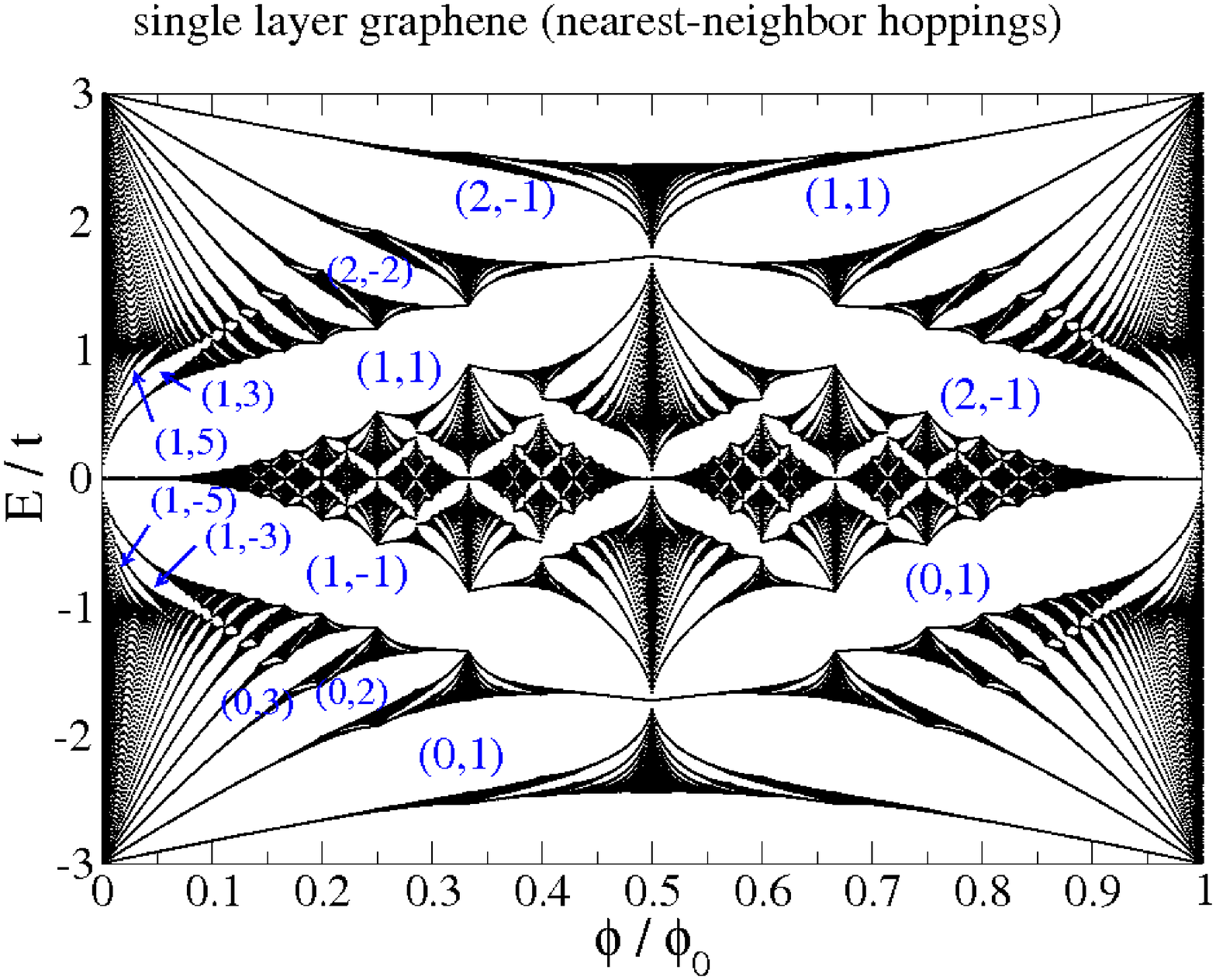} %
\includegraphics[width=0.45\textwidth]{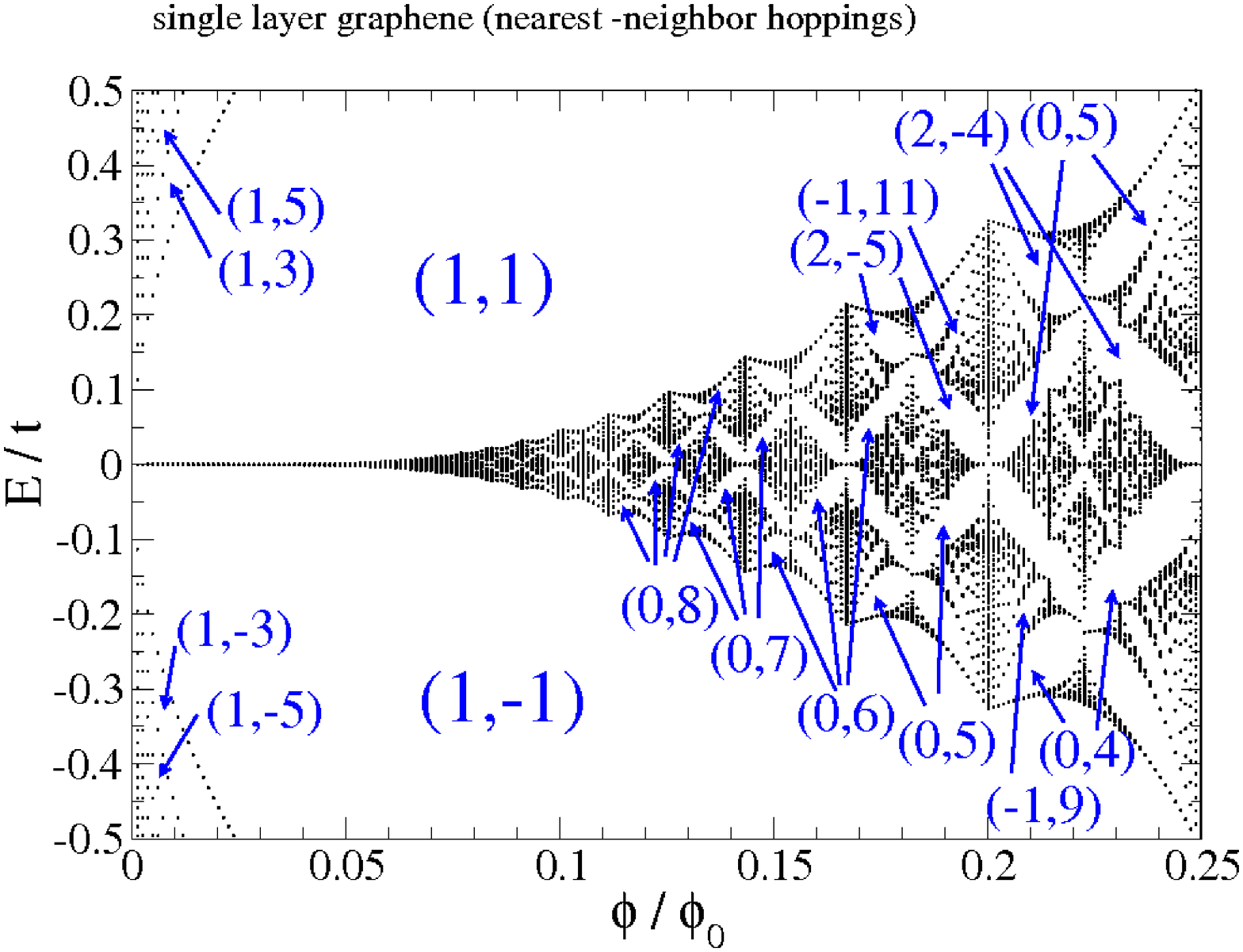} %
\end{center}
\caption{(color online).
Energy spectrum in single layer graphene
with only nearest-neighbor hoppings.
Numbers in the figures are $(s_r, t_r)$. 
Quantized value of Hall conductance is given by $t_r$.
See Eqs.~(\ref{eqDiophantine}) and (\ref{eqQHE}).
}
\label{figmono}
\end{figure}

\begin{figure}[bt]
%
\begin{center}
\includegraphics[width=0.45\textwidth]{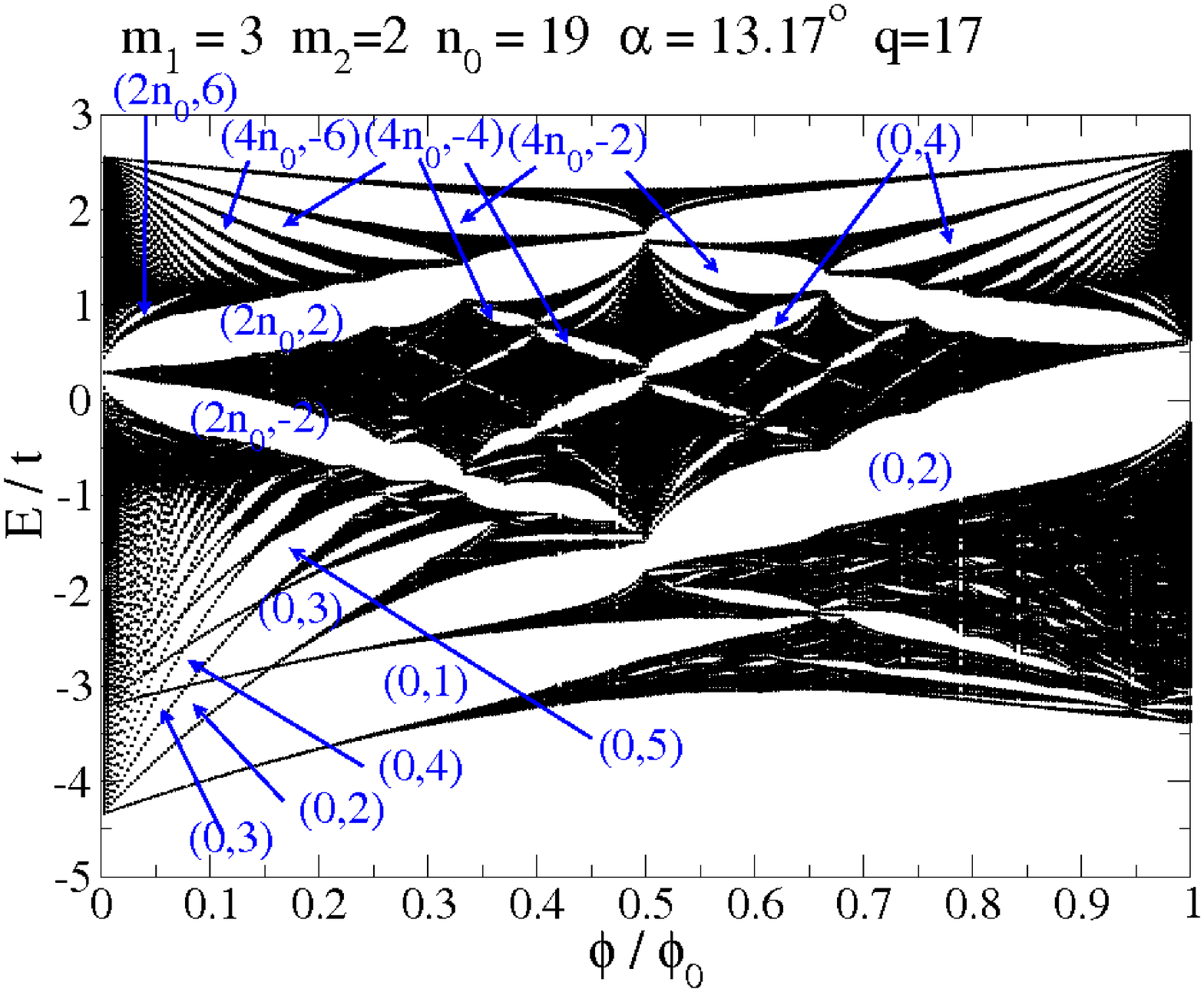} %
\includegraphics[width=0.45\textwidth]{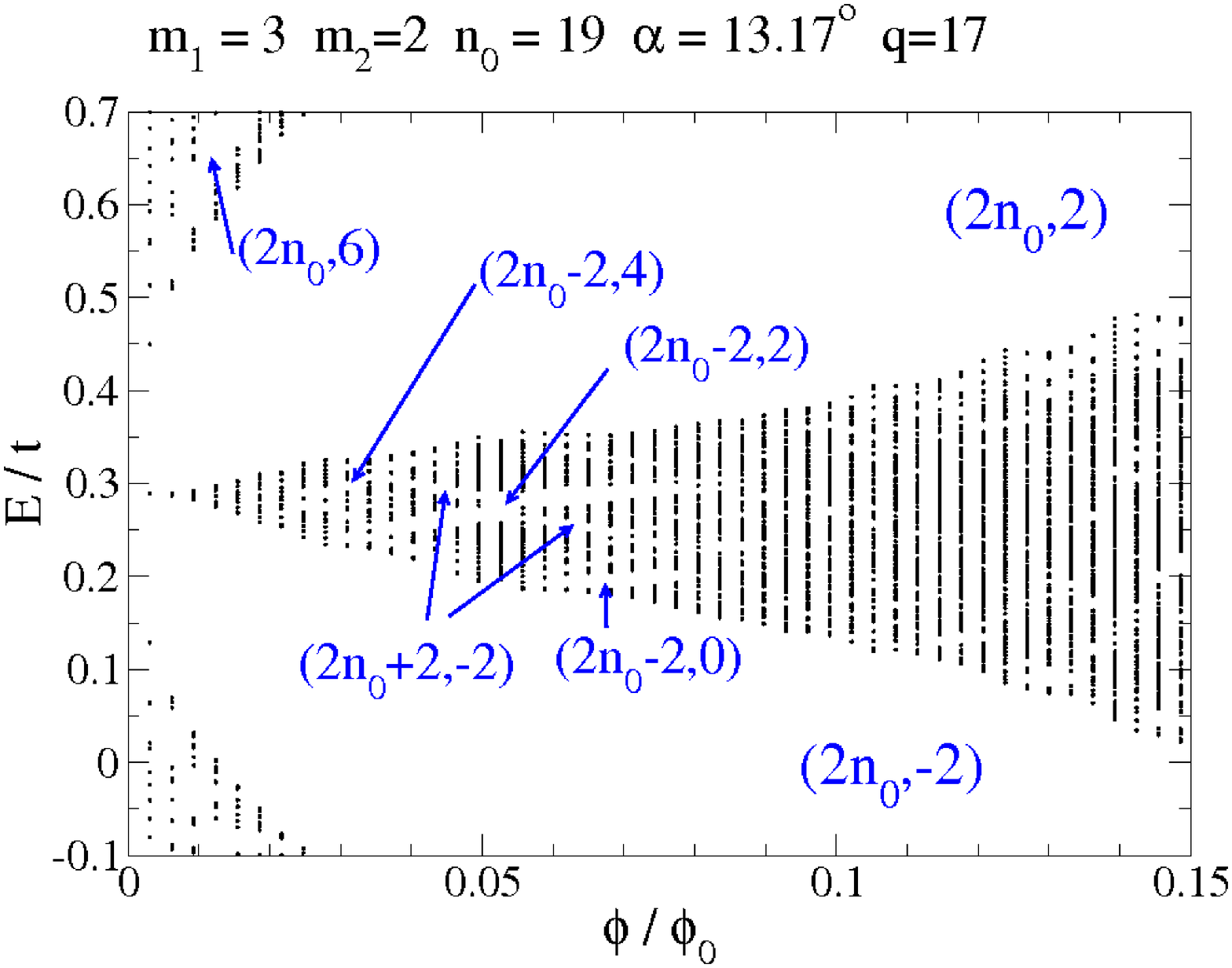} %
\end{center}
\caption{(color online).
Energy spectrum in twisted bilayer graphene 
with $(m_1, m_2) = (3,2)$.
Numbers in the figures are $(s_r, t_r)$. 
Quantized value of Hall conductance is given by $t_r$.
}
\label{figfighall32}
\end{figure}

\begin{figure}[bt]
%
\begin{center}
\includegraphics[width=0.45\textwidth]{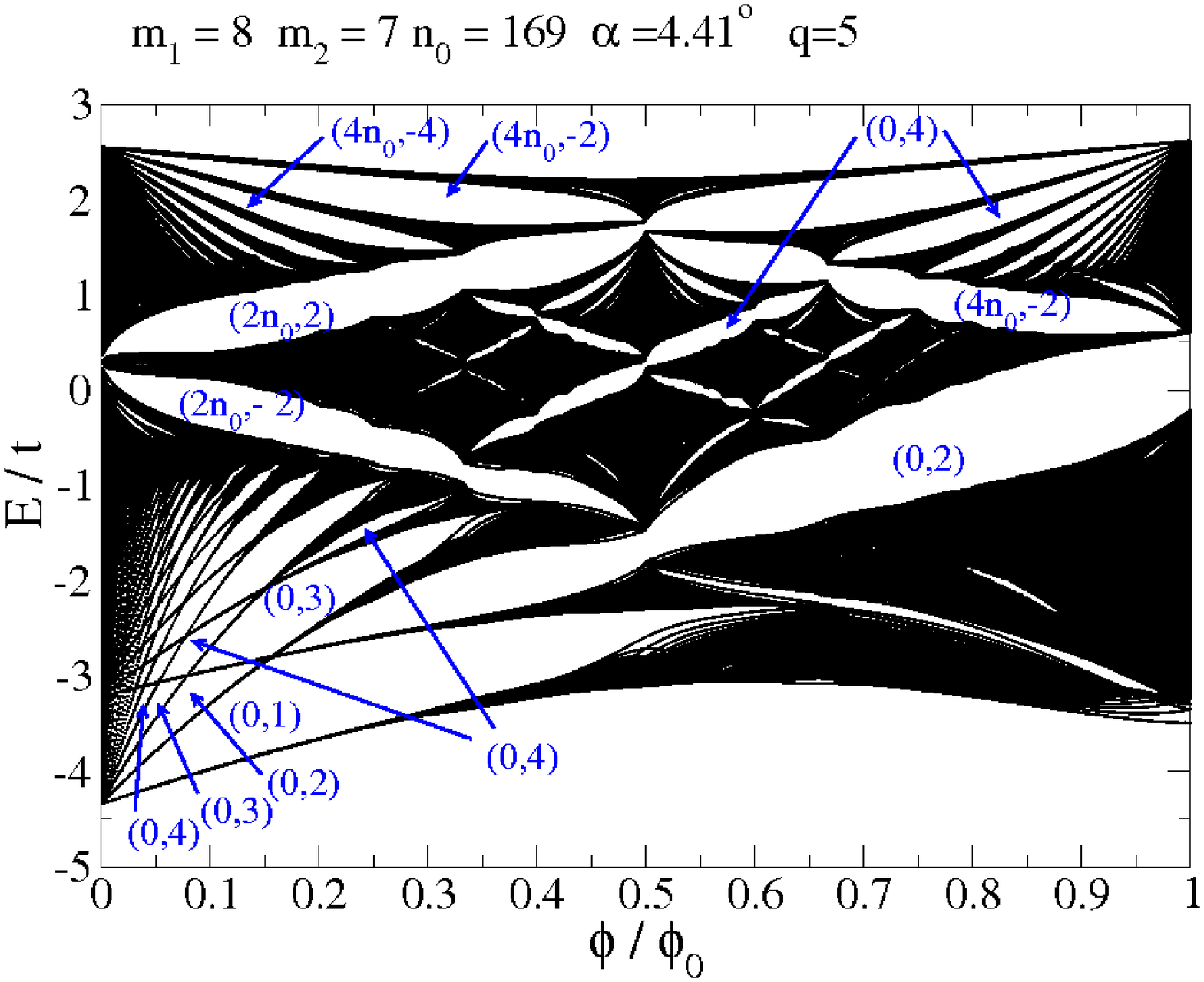} %
\includegraphics[width=0.45\textwidth]{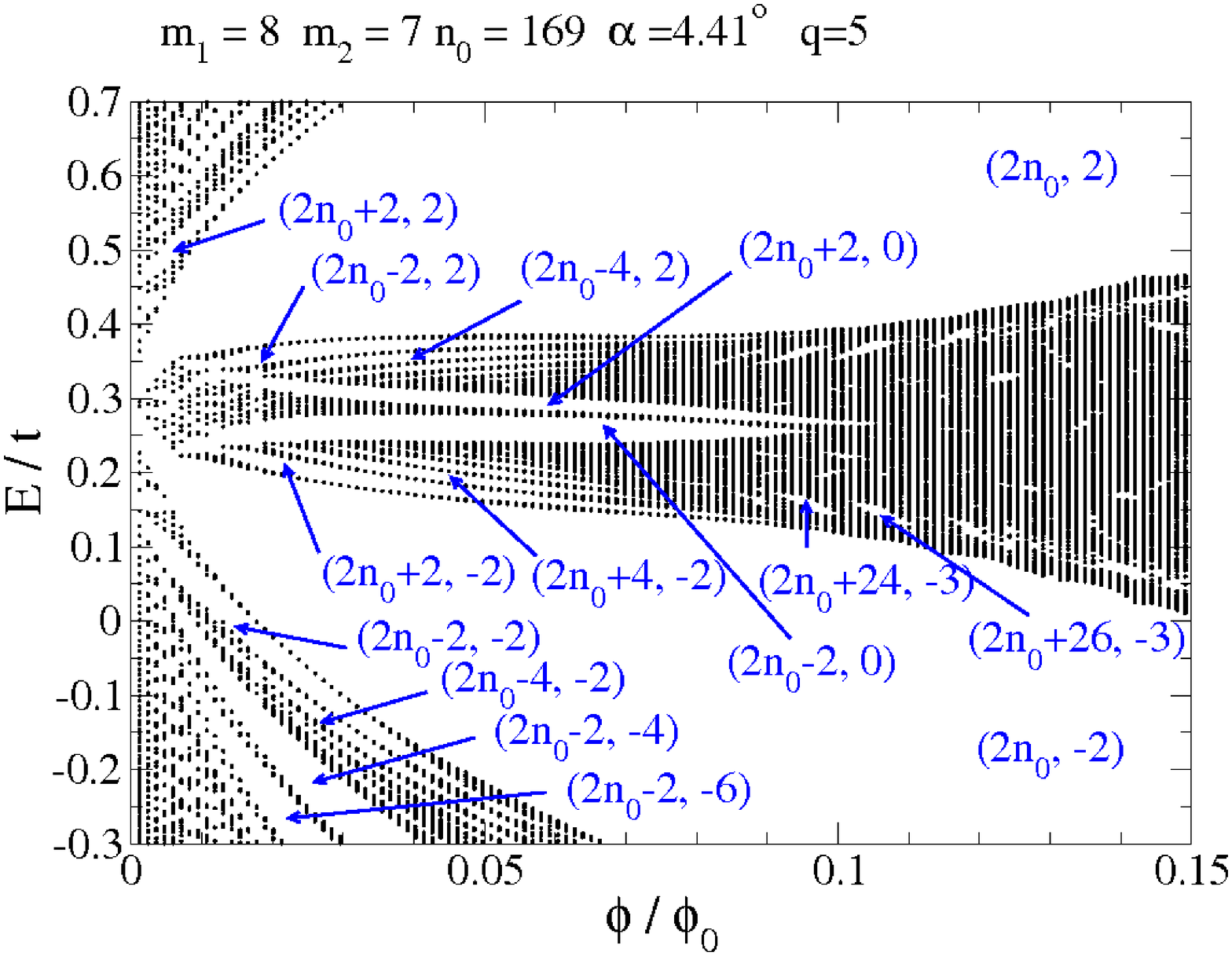} %
\end{center}
\caption{(color online).
Energy spectrum in twisted bilayer graphene with $(m_1, m_2)=(8,7)$.
Numbers in the figures are $(s_r, t_r)$. 
Quantized value of Hall conductance is given by $t_r$.
}
\label{figfighall87}
\end{figure}

\begin{figure}[bt]
%
\begin{center}
\includegraphics[width=0.45\textwidth]{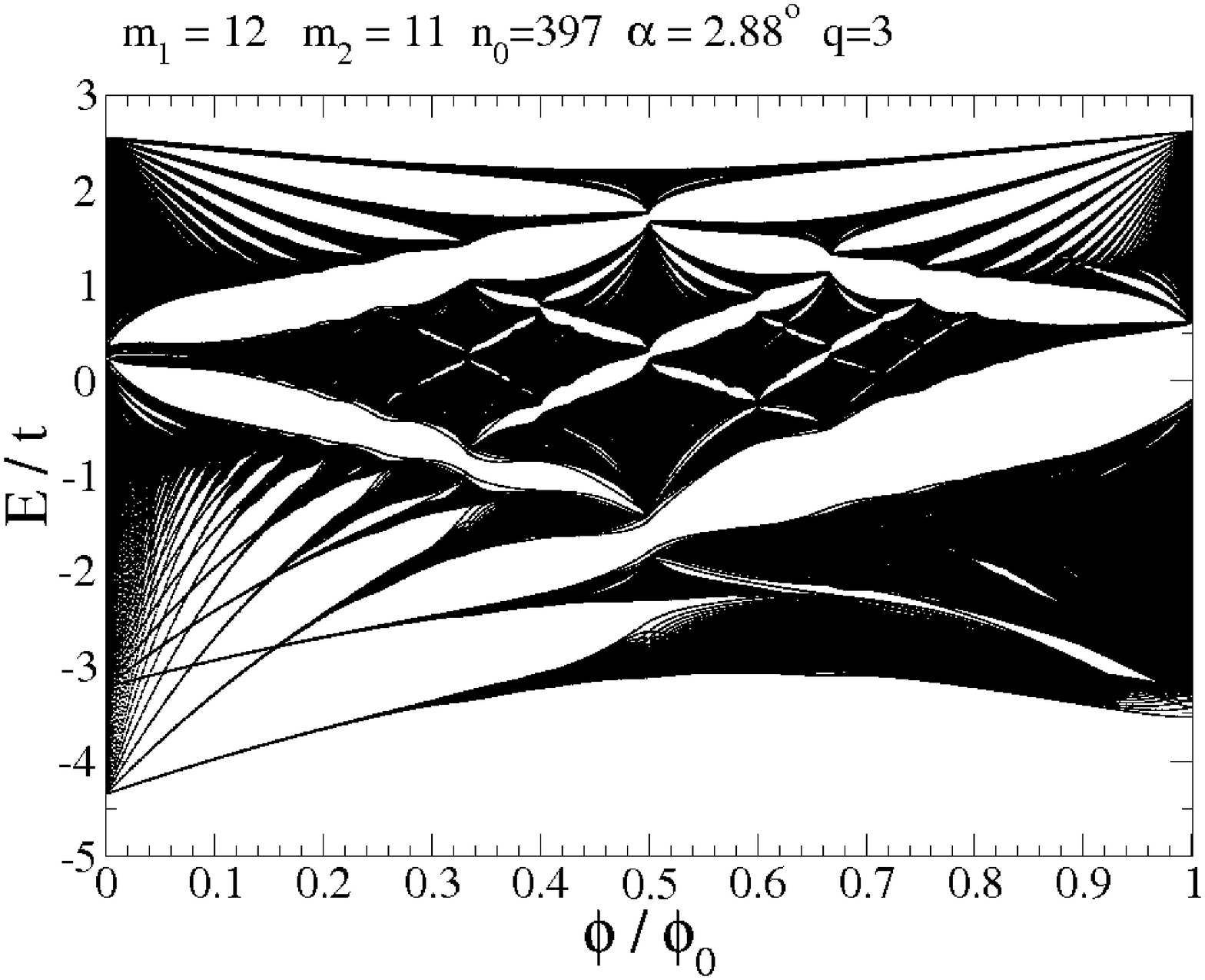} %
\includegraphics[width=0.45\textwidth]{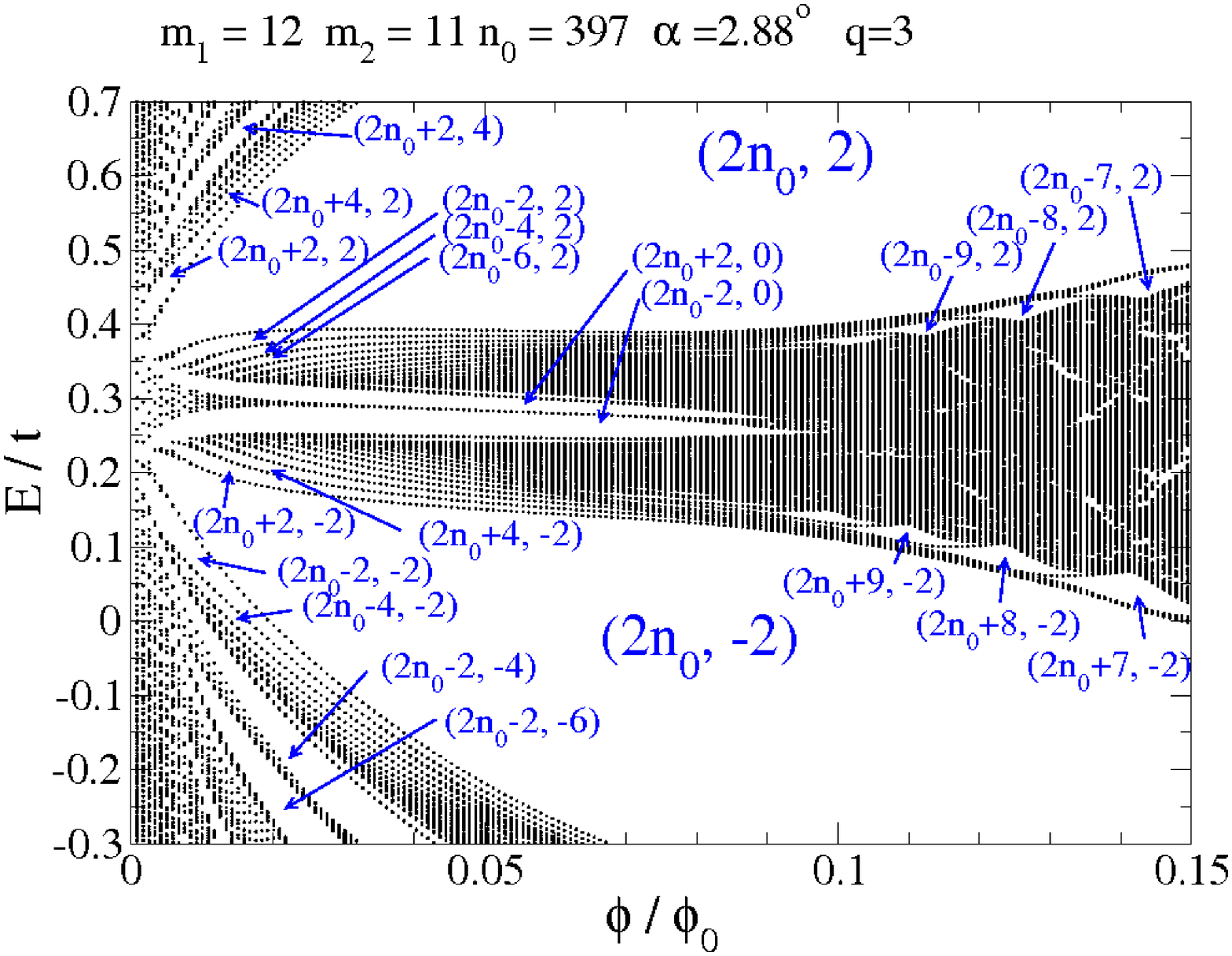} %
\end{center}
\caption{(color online).
Energy spectrum in twisted bilayer graphene
with $(m_1, m_2)=(12,11)$.
Numbers in the figures are $(s_r, t_r)$. 
Quantized value of Hall conductance is given by $t_r$.
}
\label{figfighall1211}
\end{figure}

In Figs.~\ref{figmono}, 
we plot the energy versus magnetic flux through a unit cell
in single layer graphene with
only nearest-neighbor hoppings.
In Figs.~\ref{figfighall32},  \ref{figfighall87},
 and \ref{figfighall1211},
we take parameters for the bilayer graphene\cite{Moon2012}
$t=2.7$~eV, $t_{12}=-0.48$~eV ($t_{12}/t=-0.18$), 
$|\boldsymbol{\delta}_a|=0.142$~nm, 
$d_0=0.335$~nm, and $\delta=\delta_{12}=0.0453$~nm 
($\delta/|\boldsymbol{\delta}_a|=0.184$), and
we plot the energy versus magnetic flux through a unit cell
in each layer ($\phi =\Phi/n_0$) in twisted bilayer graphene.
%
%
%
%
%
\section{Diophantine equation}
Consider the case
\begin{equation}
 \frac{\Phi}{\phi_0} = \frac{p}{q},
\end{equation}
where $\Phi$ is the flux through a supercell, $p$ and $q$ are integers.
In square lattice and honeycomb lattice
$\Phi=\phi$, and in twisted bilayer graphene $\Phi=n_0 \phi$,
where $\phi$ is the flux through a unit cell in each layer.
When  chemical potential is in the $r$th gap from the bottom, 
we have the Diophantine
equation\cite{Thouless1982,Kohmoto1985,Kohmoto1989,Sato2008}
\begin{equation}
 r = q s_r + p t_r,
\label{eqDiophantine}
\end{equation} 
which gives quantized Hall conductance by
\begin{equation}
 \sigma_{xy} = \frac{e^2}{h} t_r.
\label{eqQHE}
\end{equation}
If we take account of the spin and neglect the Zeeman energy, the
Hall conductance is multiplied by $2$.
 
\begin{figure}[bt]
%
\begin{center}
\includegraphics[width=0.45\textwidth]{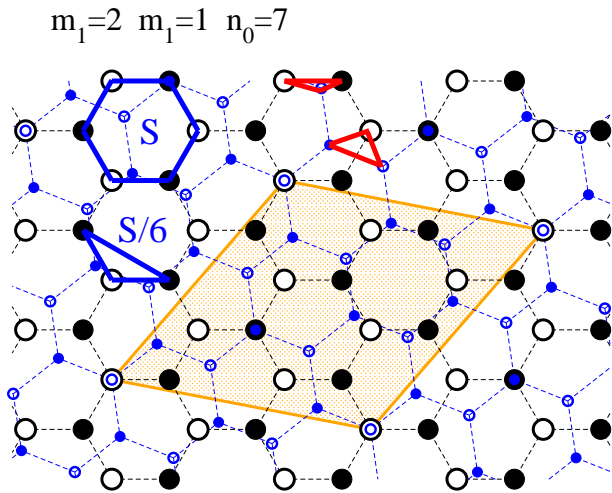} %
\end{center}
\caption{(color online).
Twisted bilayer graphene. When there are only nearest-neighbor hoppings
in each layer, an area enclosed by hoppings is $S$ (blue hexagon). 
When there are second-nearest-neighbor 
or third-nearest-neighbor hoppings in each layer, 
areas enclosed by hoppings are 
$S/6$ and its multiples (blue triangle).
 The hoppings between layers make the area enclosed by hoppings
to be irrational number times $S$ (red triangles).
}
\label{figfigtwistbl}
\end{figure}

For the tight binding
models
with only nearest-neighbor hoppings
in square lattice or honeycomb lattice, 
the flux quantum $\phi_0$ through a unit cell is 
equivalent to zero magnetic flux. 
 As a result, the energy spectrum is
 periodic with respect to 
$\phi$ with a period $\phi_0$. 
Even when we consider the models 
with long range hoppings, the energy spectra
 are periodic function of
$\phi$ with a period $2 \phi_0$ or $6 \phi_0$ 
in the square lattice or the honeycomb lattice,
respectively. 
This is because the smallest areas enclosed by hoppings are 
$1/2$ and $1/6$  of the areas of a unit cell in the square lattice
and the honeycomb lattice, respectively. See Fig.~\ref{figfigtwistbl}
for the honeycomb lattice.
The energy spectrum is also periodic with respect to $\phi$ with a
period $6 \phi_0$ for Bernal stacked bilayer graphene.
The situation is drastically changed in twisted bilayer graphene.
When there are hoppings between layers in twisted bilayer graphene, 
 projected areas enclosed by hoppings 
have irrational values as shown in the red triangles in 
Fig.~\ref{figfigtwistbl}.
As a result the energies are not periodic in $\phi$.
%

In single layer graphene, there are $2 q$ band
when flux through a supercell is
$(p/q) \phi_0$.
In Fig.~\ref{figmono}, we show $(s_r, t_r)$ for several gaps  
for single layer graphene\cite{Hasegawa2006}.
Large gaps have indices $t_r=0$, $1$ , and $2$.
The gaps, which are focused at the bottom of the band at $\phi \to 0$,
have $s_r=0$, and $t_r=1,2,3, \cdots$.
They correspond to the 
usual Landau levels. 
The gaps, which are focused at the top of the band at $\phi \to 0$,
have $s_r=2$, and $t_r=-1, -2, -3, \cdots$. 
The gaps near half filling ($E \approx 0$) and 
$\phi \ll \phi_0$ have $s_r=1$ and
$t_r = \pm 1, \pm 3, \pm 5, \cdots$, which have been 
observed in graphene\cite{Novoselov2005,Zhang2005}.
For the finite $\phi$ the band near $E \approx 0$ becomes broadened
gradually and many gaps can be seen in Fig.~\ref{figmono}.
Note that $\phi=\phi_0$ in a unit cell 
corresponds to $40,000$~T, which is not attainable 
in a present day laboratory.

\begin{figure}[bt]
\vspace{0.5cm} 
\begin{center}
\includegraphics[width=0.45\textwidth]{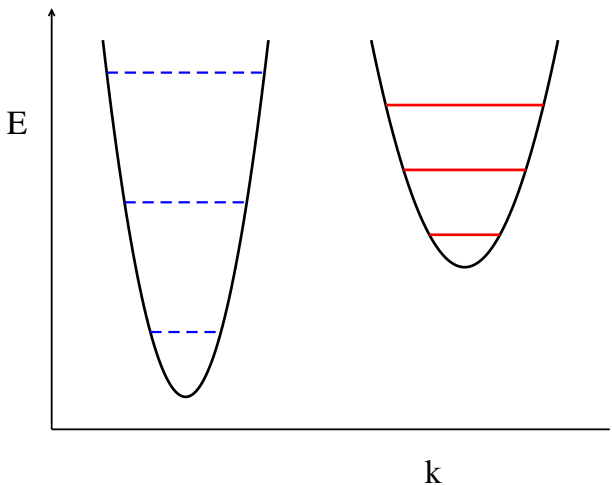} %
\end{center}
\caption{(color online).
Schematic band structure near the bottom of the energy.
There are two local minimums and  
 nearly independent Landau Levels 
(dashed blue lines and solid red lines). 
}
\label{figtwobands}
\end{figure}

In Figs.~\ref{figfighall32},~\ref{figfighall87}, and \ref{figfighall1211},
we plot the Hofstadter butterfly diagrams for twisted bilayer graphene
with $(m_1,m_2)=(3,2)$, $(8,7)$ and $(12,11)$, respectively.
Although energy gaps near the bottom have $s_r=0$ and $t_r=1,2,3, \cdots$,
for all three cases
as in single layer graphene,  
there are crossings of the bands near the bottom of
 the energy.  For example,
the gap indexed by $(0,2)$ vanishes 
at $\phi/\phi_0 \approx 0.16$ and $E/t \approx -2.8$, at which band-crossing
occurs. These crossings of bands can be understood 
by the independent Landau levels for the two local minimums
of the energy in the absence of a magnetic field (see Fig.~\ref{figtwobands}).
%
Near the top of the energy the large gaps are indexed by
$(s_r, t_r)=(4n_0,-2), (4n_0,-4), (4n_0,-6),\cdots$,
which can be understood by the fact that there are $4n_0$ bands and
nearly degenerate two local maxima of 
the energy in the absence of a magnetic field.

A very interesting feature is seen near half filling.
Besides the large gaps of $(s_r,t_r)=(2 n_0,\pm 2)$, many new gaps
become visible 
as $\alpha$ becomes small. For example,
 $(s_r,t_r)=(2 n_0 \pm 2,0), (2n_0 \pm 2,\pm 2), (2n_0 \pm 4,\pm 2),
(2n_0 \pm 6,\pm 2), \cdots$ are seen in 
the lower figures in Figs.~\ref{figfighall87} and \ref{figfighall1211}.
These new gaps are caused by a large supercell, which has $4n_0$ sites.
Since Hall conductance is given by $t_r$, the band between
the gaps with same $t_r$ ($(2 n_0, 2)$ and $(2 n_0 -2, 2)$, for example)
does not contribute to the Hall conductance.
Mathematically, that band is a Cantor set and
 consists of narrower bands and much smaller gaps. Each narrow band
gives a finite Hall conductance and the total contribution vanishes. 

\section{summary}
We obtain the Hofstadter butterfly diagram for
twisted bilayer graphene. The use of the periodic Landau gauge is
crucial. 
Due to  large number of sites ($4n_0$) 
in a supercell,
a rich structure of the Hofstadter butterfly diagram appears, 
especially near half-filling and for small rotation angle $\alpha$.
The gaps are indexed by two integers
 $s_r$ and $t_r$ (Eq.~(\ref{eqDiophantine})).
The Hall conductance is given by
$\sigma_{xy}=(e^2/h) t_r$.
While gaps with $s_r=0$, 
$1$ and $2$  are large in single layer graphene,
many gaps with $s_r=2n_0, 2n_0 \pm 2, 2n_0 \pm 4, \cdots$
become large as $\alpha$ becomes small.
Since a supercell becomes large as $\alpha \to 0$,
flux per supercell versus the flux quantum ($\Phi/\phi_0=p/q$) can be
a rational number with a small denominator $q$ 
in not an extremely strong magnetic field.
For $m_1=12$ and $m_2=11$, we obtain $n_0=392$. In that case
$100$ T corresponds to $\Phi/\phi_0\approx 1$, 
 which may be attained in experiment.

Near half-filling, there are many narrow bands, which do not 
contribute to the Hall conductance when they are completely filled.
For example, the bands between the gaps with 
$(s_r, t_r)=(2 n_0+2,2)$, $(2 n_0,2)$, $(2n_0-2,2)$, $(2n_0-4,2)$, etc. 
at $E/t \approx 0.3 \sim 0.4$
in the lower figure in Fig.~\ref{figfighall1211} do not change the 
quantized value of the Hall conductance $t_r=2$.  
Similarly the band between the gaps with 
$(s_r, t_r)=(2 n_0+2,0)$ and $(2 n_0-2,0)$ at $E/t \approx 0.3$
in the lower figure in Fig.~\ref{figfighall1211}
does not change the 
quantized value of the Hall conductance $t_r=0$. 
The narrow bands, in fact, consist of even narrower bands, since
an energy spectrum is a Cantor set. 
One can expect other value of $t_r$
in the narrower band.
It may be possible to observe these phenomena experimentally.

\appendix
\section{periodic Landau gauge in square lattice}
\label{labAppA}
\begin{figure}[b]
\begin{center}
\vspace{1.0cm} 
\includegraphics[width=0.4\textwidth]{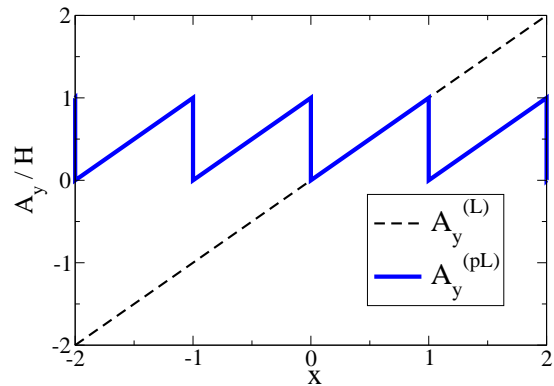} %
\end{center}
\caption{(color online).
The $y$ component of vector potentials for Landau gauge ($A^{(L)}_y$,
dashed line) and periodic Landau gauge 
($A^{(pL)}_y$, thick blue line) for the square lattice.
}
\label{figAy}
\end{figure}
\begin{figure}[b]
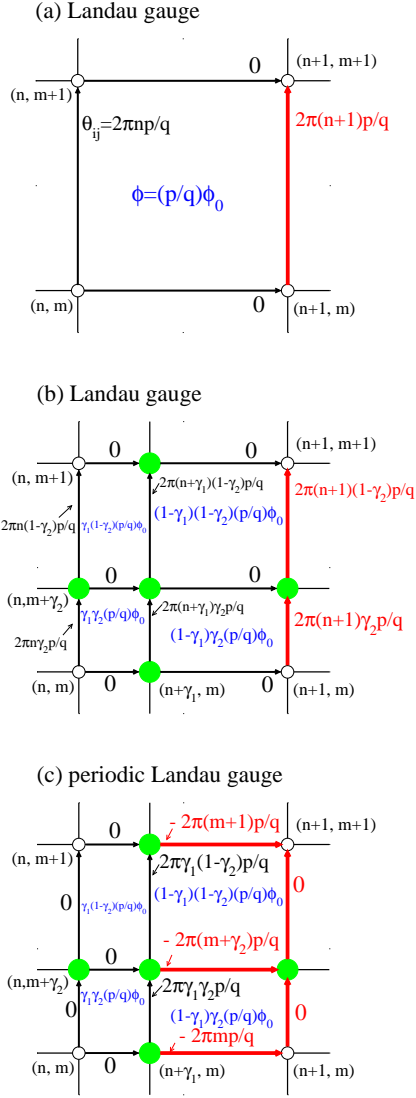

%
\begin{center}
\includegraphics[width=0.332\textwidth]{fig9a.eps} 
\includegraphics[width=0.33\textwidth]{fig9b.eps}  
\includegraphics[width=0.33\textwidth]{fig9c.eps} %
\end{center}
\caption{(color online).
(a) The numbers beside the arrows are 
the phases in the usual Landau gauge
for the square lattice 
with flux $(p/q)\phi_0$ in a unit cell.
The system is periodic in the $x$ direction with period $q$.
(b) Sites are added at $(n+\gamma_1,m)$, $(n,m+\gamma_2)$, and
$(n+\gamma_1,m+\gamma_2)$, where $n$ and $m$ are integers, 
$0< \gamma_1 <1$ and $0< \gamma_2 <1$
(filled green circles).
Blue letters are flux in rectangles.
The system is periodic only when 
$\gamma_2$ is a rational number.
(c) The phases  (beside arrows)
and the flux through rectangles (blue letters)
in the periodic Landau gauge.
The system is periodic in the $x$ direction with period $1$ and in 
the $y$ direction
with period $q$. 
}
\label{figphase}
\end{figure}
We explain the periodic Landau gauge
in the square lattice.
A vector potential $\mathbf{A}$ gives a magnetic field
\begin{equation}
\mathbf{H} = \nabla \times \mathbf{A}.
\end{equation}
For a uniform magnetic field $\mathbf{H} = H \hat{\mathbf{z}}$,
one can take the Landau gauge
\begin{equation}
 \mathbf{A}^{(L)} = H x \hat{\mathbf{y}},
\label{eqeqLandau}
\end{equation}
where $\hat{\mathbf{y}}$ is the unit vector along the $y$ direction
and there is no dependence on $y$.
In this case, however, $A^{(L)}_y$ 
is not periodic in the $x$ direction as shown 
by the dashed line
in Fig.~\ref{figAy}.
We can obtain other vector potential 
by gauge transformation, i.e., 
adding $\nabla \chi(\mathbf{r})$ to $\mathbf{A}$.
It is crucial 
to have periodicity in a gauge of twisted bilayer graphene.
 We take 
\begin{equation}
\chi(\mathbf{r}) = -H \lfloor x \rfloor y,
\end{equation}
and so, 
\begin{align}
 \mathbf{A}^{(pL)} &= H \left( x \hat{\mathbf{y}} - \nabla 
(\lfloor x \rfloor y) \right) \nonumber \\
 &= H \left( (x-\lfloor x \rfloor ) \hat{\mathbf{y}}
 - y \sum_{n=-\infty}^{\infty}\delta(x-n+\epsilon)\hat{\mathbf{x}}
\right),
\label{eqA3}
\end{align}
where 
$\epsilon$ is an infinitesimal and
$\lfloor x \rfloor $ is the floor function (largest integer 
not greater than $x$), 
i.e. $x-\lfloor x \rfloor$ is the fractional part of $x$.
In this gauge, which we call the periodic Landau gauge, 
$A^{(pL)}_y$ is  periodic with respect to $x$ with period 1, 
as shown in Fig.~\ref{figAy}.
In order to make $A^{(pL)}_y$ be periodic in the $y$ direction
$A^{(pL)}_y$ is a discontinuous function of $x$ as shown in Fig.~\ref{figAy}
and $A^{(pL)}_x$ is the sum of delta functions. 
These singular functions do not cause any problems.
We have no ambiguity in the phase factor $\theta_{ij}$,
since we have added the infinitesimal $\epsilon$\cite{Trellakis2003,Nemec2007}.  

 Note that $A^{(pL)}_x$ depends on $y$
and it is not periodic in the $y$ direction.
However, the dependence of $y$ in $\mathbf{A}^{(pL)}$ 
appears always with the delta function,
so  $\exp(i \theta_{ij})$
is periodic in the $y$ direction,
as we show below.

The difference of the usual Landau gauge and the 
periodic Landau gauge is seen in Fig.~\ref{figphase}
in which a uniform magnetic field 
with $(p/q)\phi_0$ through a unit cell is applied
to square lattice.
If we take the usual Landau gauge, the phase factor 
is zero except vertical links as shown in Fig.~\ref{figphase} (a).
The phase factor for the vertical links at $x=n$ is $\theta_{ij}=2\pi n p/q$.
The periodicity in the $x$ direction is $q$ times larger than the  
periodicity in the absence of a magnetic field. 
However, if there are other sites in a unit cell,
the periodicity of the system is changed 
(this is the case in twisted bilayer graphene, where there are $4 n_0$
sites in the supercell (see Figs.~\ref{figfig21} and \ref{figfigtwistbl})).
In order to demonstrate it in the square lattice,
we add sites  at $(n+\gamma_1,m)$, $(n,m+\gamma_2)$, and
at $(n+\gamma_1,m+\gamma_2)$, 
where $n$ and $m$ are integers, $0 < \gamma_1 < 1$ and
$0 < \gamma_2 < 1$, 
as shown by the filled green circles in Fig.~\ref{figphase}(b).
The phase factors for the links connecting 
neighbor 
sites are 
shown in Fig.~\ref{figphase}(b). 
The phase factor for the vertical link between
$\mathbf{r}_i=(n,m)$ and $\mathbf{r}_j=(n,m+\gamma_2)$ 
is $\theta_{ij}=2 \pi n \gamma_2 p/q$. 
If $\gamma_2$ is an irrational number, 
$\exp(i \theta_{ij})$ cannot be periodic with respect to $x$.

The periodicity is recovered by taking the periodic Landau gauge
(Eq.~(\ref{eqA3})).
The periodicity is 1 in the $x$ direction, since $\mathbf{A}^{(pL)}$
is periodic in the $x$ direction.
The delta functions in Eq.~(\ref{eqA3}) make
the nonzero phases for the horizontal links 
as shown by thick red horizontal arrows in Fig,~\ref{figphase}(c). 
The magnetic flux through each small rectangles is obtained by the
sum of the surrounding phases $\theta_{ij}$ and it is
proportional to the area.
The phase factor for the horizontal link between
$\mathbf{r}_i=(n+\gamma_1,m)$ and $\mathbf{r}_j=(n+1,m)$
is $\theta_{ij}=-2\pi m p/q$,
which does not depend on $\gamma_1$.
The periodicity 
in the $y$ direction is $q$ times larger than that
without magnetic field in the periodic Landau gauge.
\section{periodic Landau gauge in non-square lattice}
\label{labAppB}
%
\begin{figure}[b]
%
\begin{center}
\vspace{0.5cm}
\includegraphics[width=0.45\textwidth]{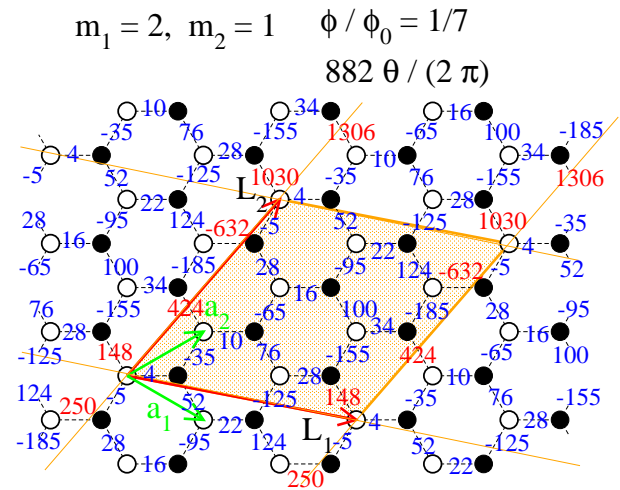} 
\end{center}
\caption{(color online).
The first layer of 
twisted bilayer graphene with $(m_1,m_2)=(2,1)$ ($n_0=7$).
The numbers on the links between the nearest neighbor sites 
show the phase $\theta_{ij}$  
in the first layer
in units of $2\pi/(18 n_0^2) =2 \pi/882$ when 
 the flux per supercell is $\phi_0$, i.e. the flux per unit hexagon is 
$\phi_0/n_0$
(We obtain
$(4-(-35)+424-34-(-155)-148)\times 2\pi/882=2\pi/7$, for example).
We take
the periodic Landau gauge (Eq.~(\ref{eqApLnsq})).
The contribution from the delta function
in Eq.~(\ref{eqApLnsq}) is finite for the red numbers, while 
it is zero for the blue numbers.
}
\label{figfig21h}
\end{figure}
The periodic Landau gauge discussed
 in Appendix~\ref{labAppA} is generalized 
to  non-square two-dimensional lattices, which has
primitive vectors of the supercell  
 $\mathbf{L}_1$ and $\mathbf{L_2}$,
which are not orthogonal. 
The reciprocal lattice vectors are
\begin{align}
\mathbf{F}_1 &= 2\pi \frac{\mathbf{L}_2 \times 
\hat{\mathbf{z}}}{(\mathbf{L}_1
\times \mathbf{L}_2)\cdot \hat{\mathbf{z}}} ,
\end{align}
and
\begin{align}
\mathbf{F}_2 &= 2\pi \frac{\hat{\mathbf{z}} 
\times \mathbf{L}_1}{(\mathbf{L}_1
\times \mathbf{L}_2)\cdot \hat{\mathbf{z}}} .
\end{align}
We define oblique coordinate system $(\xi_1, \xi_2)$ by
\begin{equation}
\mathbf{r}=
\left( \begin{array}{c}
x \\y \end{array} \right) =
\xi_1 \mathbf{L}_1 +\xi_2 \mathbf{L}_2,
\end{equation}
where $x$ and $y$ are the coordinates in a orthogonal system.
For twisted bilayer graphene we have
\begin{align}
 \xi_1 &= \frac{1}{n_0} \left( \frac{\sqrt{3}}{3}(m_1+2m_2) x -m_1 y\right),\\
 \xi_2 &= \frac{1}{n_0} \left( \frac{\sqrt{3}}{3}(m_1- m_2) x +
(m_1+m_2) y\right).
\end{align}
The reciprocal vectors are
\begin{align}
\mathbf{F}_1 
&=\frac{m_1+m_2}{n_0} \mathbf{G}_1 + \frac{m_2}{n_0} \mathbf{G}_2 ,
\end{align}
and
\begin{align}
\mathbf{F}_2 
&=\frac{-m_2}{n_0} \mathbf{G}_1 + \frac{m_1}{n_0} \mathbf{G}_2.
\end{align}
The Landau gauge for the oblique coordinate system is
\begin{equation}
\mathbf{A}^{(L,nsq)}=\frac{S H}{2\pi} \xi_1 \mathbf{F}_2,
\label{eqALnsq}
\end{equation}
which is a generalization of  Eq.~(\ref{eqeqLandau})
to the non-square lattice.
To make the vector potential periodic with respect to $\xi_1$,
we take the periodic Landau gauge as 
\begin{align}
\mathbf{A}^{(pL,nsq)}=&\frac{S H}{2\pi} \Bigl( (\xi_1-\lfloor \xi_1
 \rfloor) \mathbf{F}_2 \nonumber \\
&-\xi_2\sum_{n=-\infty}^{\infty} \delta (\xi_1-n+\epsilon) 
\mathbf{F}_1 \Bigr).
\label{eqApLnsq}
\end{align}
In Fig.~\ref{figfig21h} we show an example of 
$\theta_{ij}$ between the nearest neighbor sites in the 
first layer for $m_1=2$, $m_2=1$ ($n_0=7$)
and the flux through a supercell is $\phi_0$ ($p/q=1/1$).
Note that the phase factor $\theta_{ij}$ 
crossing the line $\xi_1=n$ (red numbers in Fig.~\ref{figfig21h})
is not periodic in the $\mathbf{L}_2$ direction
but $\exp(i \theta_{ij})$ is periodic.

The advantage of taking the periodic Landau gauge is that
the vector potential $\mathbf{A}$ is periodic in $\xi_1$ with 
period 1. This ensures the periodicity of the system in $\mathbf{L}_1$ 
direction. The system has a periodicity in the $\mathbf{L}_2$ direction
if the flux per supercell is an rational number times $\phi_0$.

If the flux per a supercell is 
a rational number $p/q$ with integers 
$p$ and $q$, i.e.
\begin{equation}
\Phi=n_0 \phi= S H =\frac{p}{q}\phi_0, 
\end{equation}
the phase factor $\theta_{ij}$ in Eq.~(\ref{eqtheta}) 
with the periodic Landau 
gauge (Eq.~(\ref{eqApLnsq})) has the same value 
when $\mathbf{r}_i$ and $\mathbf{r}_j$
are translated by $\mathbf{L}_1$.
It has also the same value when $\mathbf{r}_i$ and $\mathbf{r}_j$
are translated by $\mathbf{L}_2$ if the link connecting
$\mathbf{r}_i$ and $\mathbf{r}_j$ does not cross the line $\xi_1=n$,
where $n$ is an integer. 
If the link connecting
$\mathbf{r}_i$ and $\mathbf{r}_j$ crosses the line $\xi_1=n$,
$\theta_{ij}$ increases by
$\pm 2\pi p/q$ when $\mathbf{r}_i$ and $\mathbf{r}_j$
are translated by $\mathbf{L}_2$ ($\pm$ depends on the sign of $\mathbf{F}_1
\cdot d\boldsymbol{\ell}$). 
Therefore periodicity 
must be  $q$ times larger in the
$\mathbf{L}_2$ direction.

\clearpage

\bibliography{gauge}

\begin{thebibliography}{31}%
\makeatletter
\providecommand \@ifxundefined [1]{%
 \@ifx{#1\undefined}
}%
\providecommand \@ifnum [1]{%
 \ifnum #1\expandafter \@firstoftwo
 \else \expandafter \@secondoftwo
 \fi
}%
\providecommand \@ifx [1]{%
 \ifx #1\expandafter \@firstoftwo
 \else \expandafter \@secondoftwo
 \fi
}%
\providecommand \natexlab [1]{#1}%
\providecommand \enquote  [1]{``#1''}%
\providecommand \bibnamefont  [1]{#1}%
\providecommand \bibfnamefont [1]{#1}%
\providecommand \citenamefont [1]{#1}%
\providecommand \href@noop [0]{\@secondoftwo}%
\providecommand \href [0]{\begingroup \@sanitize@url \@href}%
\providecommand \@href[1]{\@@startlink{#1}\@@href}%
\providecommand \@@href[1]{\endgroup#1\@@endlink}%
\providecommand \@sanitize@url [0]{\catcode `\\12\catcode `\$12\catcode
  `\&12\catcode `\#12\catcode `\^12\catcode `\_12\catcode `\%12\relax}%
\providecommand \@@startlink[1]{}%
\providecommand \@@endlink[0]{}%
\providecommand \url  [0]{\begingroup\@sanitize@url \@url }%
\providecommand \@url [1]{\endgroup\@href {#1}{\urlprefix }}%
\providecommand \urlprefix  [0]{URL }%
\providecommand \Eprint [0]{\href }%
\providecommand \doibase [0]{http://dx.doi.org/}%
\providecommand \selectlanguage [0]{\@gobble}%
\providecommand \bibinfo  [0]{\@secondoftwo}%
\providecommand \bibfield  [0]{\@secondoftwo}%
\providecommand \translation [1]{[#1]}%
\providecommand \BibitemOpen [0]{}%
\providecommand \bibitemStop [0]{}%
\providecommand \bibitemNoStop [0]{.\EOS\space}%
\providecommand \EOS [0]{\spacefactor3000\relax}%
\providecommand \BibitemShut  [1]{\csname bibitem#1\endcsname}%
\let\auto@bib@innerbib\@empty
\bibitem [{\citenamefont {Novoselov}\ \emph {et~al.}(2004)\citenamefont
  {Novoselov}, \citenamefont {Geim}, \citenamefont {Morozov}, \citenamefont
  {Jaing}, \citenamefont {Zhang}, \citenamefont {Dubonos}, \citenamefont
  {Grigorieva},\ and\ \citenamefont {Firsov}}]{Novoselov2004}%
  \BibitemOpen
  \bibfield  {author} {\bibinfo {author} {\bibfnamefont {K.~S.}\ \bibnamefont
  {Novoselov}}, \bibinfo {author} {\bibfnamefont {A.~K.}\ \bibnamefont {Geim}},
  \bibinfo {author} {\bibfnamefont {S.~V.}\ \bibnamefont {Morozov}}, \bibinfo
  {author} {\bibfnamefont {D.}~\bibnamefont {Jaing}}, \bibinfo {author}
  {\bibfnamefont {Y.}~\bibnamefont {Zhang}}, \bibinfo {author} {\bibfnamefont
  {S.~V.}\ \bibnamefont {Dubonos}}, \bibinfo {author} {\bibfnamefont {I.~V.}\
  \bibnamefont {Grigorieva}}, \ and\ \bibinfo {author} {\bibfnamefont {A.~A.}\
  \bibnamefont {Firsov}},\ }\href@noop {} {\bibfield  {journal} {\bibinfo
  {journal} {Science}\ }\textbf {\bibinfo {volume} {306}},\ \bibinfo {pages}
  {666} (\bibinfo {year} {2004})}\BibitemShut {NoStop}%
\bibitem [{\citenamefont {McCann}\ and\ \citenamefont
  {Fal'ko}(2006)}]{McCann2006PRL}%
  \BibitemOpen
  \bibfield  {author} {\bibinfo {author} {\bibfnamefont {E.}~\bibnamefont
  {McCann}}\ and\ \bibinfo {author} {\bibfnamefont {V.~I.}\ \bibnamefont
  {Fal'ko}},\ }\href {\doibase 10.1103/PhysRevLett.96.086805} {\bibfield
  {journal} {\bibinfo  {journal} {Phys. Rev. Lett.}\ }\textbf {\bibinfo
  {volume} {96}},\ \bibinfo {pages} {086805} (\bibinfo {year}
  {2006})}\BibitemShut {NoStop}%
\bibitem [{\citenamefont {Shallcross}\ \emph {et~al.}(2010)\citenamefont
  {Shallcross}, \citenamefont {Sharma}, \citenamefont {Kandelaki},\ and\
  \citenamefont {Pankratov}}]{Shallcross2010}%
  \BibitemOpen
  \bibfield  {author} {\bibinfo {author} {\bibfnamefont {S.}~\bibnamefont
  {Shallcross}}, \bibinfo {author} {\bibfnamefont {S.}~\bibnamefont {Sharma}},
  \bibinfo {author} {\bibfnamefont {E.}~\bibnamefont {Kandelaki}}, \ and\
  \bibinfo {author} {\bibfnamefont {O.~A.}\ \bibnamefont {Pankratov}},\ }\href
  {\doibase 10.1103/PhysRevB.81.165105} {\bibfield  {journal} {\bibinfo
  {journal} {Phys. Rev. B}\ }\textbf {\bibinfo {volume} {81}},\ \bibinfo
  {pages} {165105} (\bibinfo {year} {2010})}\BibitemShut {NoStop}%
\bibitem [{\citenamefont {Novoselov}\ \emph {et~al.}(2005)\citenamefont
  {Novoselov}, \citenamefont {Geim}, \citenamefont {Morozov}, \citenamefont
  {Jiang}, \citenamefont {Katsnelson}, \citenamefont {Grigorieva},
  \citenamefont {Dubonos},\ and\ \citenamefont {Firsov}}]{Novoselov2005}%
  \BibitemOpen
  \bibfield  {author} {\bibinfo {author} {\bibfnamefont {K.~S.}\ \bibnamefont
  {Novoselov}}, \bibinfo {author} {\bibfnamefont {A.~K.}\ \bibnamefont {Geim}},
  \bibinfo {author} {\bibfnamefont {S.~V.}\ \bibnamefont {Morozov}}, \bibinfo
  {author} {\bibfnamefont {D.}~\bibnamefont {Jiang}}, \bibinfo {author}
  {\bibfnamefont {M.~I.}\ \bibnamefont {Katsnelson}}, \bibinfo {author}
  {\bibfnamefont {I.~V.}\ \bibnamefont {Grigorieva}}, \bibinfo {author}
  {\bibfnamefont {S.~V.}\ \bibnamefont {Dubonos}}, \ and\ \bibinfo {author}
  {\bibfnamefont {A.~A.}\ \bibnamefont {Firsov}},\ }\href {\doibase
  10.1038/nature04233} {\bibfield  {journal} {\bibinfo  {journal} {Nature}\
  }\textbf {\bibinfo {volume} {438}},\ \bibinfo {pages} {197 } (\bibinfo {year}
  {2005})}\BibitemShut {NoStop}%
\bibitem [{\citenamefont {Zhang}\ \emph {et~al.}(2005)\citenamefont {Zhang},
  \citenamefont {Tan}, \citenamefont {Stormer},\ and\ \citenamefont
  {Kim}}]{Zhang2005}%
  \BibitemOpen
  \bibfield  {author} {\bibinfo {author} {\bibfnamefont {Y.}~\bibnamefont
  {Zhang}}, \bibinfo {author} {\bibfnamefont {Y.-W.}\ \bibnamefont {Tan}},
  \bibinfo {author} {\bibfnamefont {H.~L.}\ \bibnamefont {Stormer}}, \ and\
  \bibinfo {author} {\bibfnamefont {P.}~\bibnamefont {Kim}},\ }\href {\doibase
  10.1038/nature04235} {\bibfield  {journal} {\bibinfo  {journal} {Nature}\
  }\textbf {\bibinfo {volume} {438}},\ \bibinfo {pages} {201 } (\bibinfo {year}
  {2005})}\BibitemShut {NoStop}%
\bibitem [{\citenamefont {Hasegawa}\ and\ \citenamefont
  {Kohmoto}(2006)}]{Hasegawa2006}%
  \BibitemOpen
  \bibfield  {author} {\bibinfo {author} {\bibfnamefont {Y.}~\bibnamefont
  {Hasegawa}}\ and\ \bibinfo {author} {\bibfnamefont {M.}~\bibnamefont
  {Kohmoto}},\ }\href {\doibase 10.1103/PhysRevB.74.155415} {\bibfield
  {journal} {\bibinfo  {journal} {Phys. Rev. B}\ }\textbf {\bibinfo {volume}
  {74}},\ \bibinfo {pages} {155415} (\bibinfo {year} {2006})}\BibitemShut
  {NoStop}%
\bibitem [{\citenamefont {Hatsugai}\ \emph {et~al.}(2006)\citenamefont
  {Hatsugai}, \citenamefont {Fukui},\ and\ \citenamefont
  {Aoki}}]{Hatsugai2006}%
  \BibitemOpen
  \bibfield  {author} {\bibinfo {author} {\bibfnamefont {Y.}~\bibnamefont
  {Hatsugai}}, \bibinfo {author} {\bibfnamefont {T.}~\bibnamefont {Fukui}}, \
  and\ \bibinfo {author} {\bibfnamefont {H.}~\bibnamefont {Aoki}},\ }\href
  {\doibase 10.1103/PhysRevB.74.205414} {\bibfield  {journal} {\bibinfo
  {journal} {Phys. Rev. B}\ }\textbf {\bibinfo {volume} {74}},\ \bibinfo
  {pages} {205414} (\bibinfo {year} {2006})}\BibitemShut {NoStop}%
\bibitem [{\citenamefont {Dietl}\ \emph {et~al.}(2008)\citenamefont {Dietl},
  \citenamefont {Pi\'echon},\ and\ \citenamefont {Montambaux}}]{Dietl2008}%
  \BibitemOpen
  \bibfield  {author} {\bibinfo {author} {\bibfnamefont {P.}~\bibnamefont
  {Dietl}}, \bibinfo {author} {\bibfnamefont {F.}~\bibnamefont {Pi\'echon}}, \
  and\ \bibinfo {author} {\bibfnamefont {G.}~\bibnamefont {Montambaux}},\
  }\href {\doibase 10.1103/PhysRevLett.100.236405} {\bibfield  {journal}
  {\bibinfo  {journal} {Phys. Rev. Lett.}\ }\textbf {\bibinfo {volume} {100}},\
  \bibinfo {pages} {236405} (\bibinfo {year} {2008})}\BibitemShut {NoStop}%
\bibitem [{\citenamefont {Nemec}\ and\ \citenamefont
  {Cuniberti}(2007)}]{Nemec2007}%
  \BibitemOpen
  \bibfield  {author} {\bibinfo {author} {\bibfnamefont {N.}~\bibnamefont
  {Nemec}}\ and\ \bibinfo {author} {\bibfnamefont {G.}~\bibnamefont
  {Cuniberti}},\ }\href {\doibase 10.1103/PhysRevB.75.201404} {\bibfield
  {journal} {\bibinfo  {journal} {Phys. Rev. B}\ }\textbf {\bibinfo {volume}
  {75}},\ \bibinfo {pages} {201404} (\bibinfo {year} {2007})}\BibitemShut
  {NoStop}%
\bibitem [{\citenamefont {Thouless}\ \emph {et~al.}(1982)\citenamefont
  {Thouless}, \citenamefont {Kohmoto}, \citenamefont {Nightingale},\ and\
  \citenamefont {den Nijs}}]{Thouless1982}%
  \BibitemOpen
  \bibfield  {author} {\bibinfo {author} {\bibfnamefont {D.~J.}\ \bibnamefont
  {Thouless}}, \bibinfo {author} {\bibfnamefont {M.}~\bibnamefont {Kohmoto}},
  \bibinfo {author} {\bibfnamefont {M.~P.}\ \bibnamefont {Nightingale}}, \ and\
  \bibinfo {author} {\bibfnamefont {M.}~\bibnamefont {den Nijs}},\ }\href
  {\doibase 10.1103/PhysRevLett.49.405} {\bibfield  {journal} {\bibinfo
  {journal} {Phys. Rev. Lett.}\ }\textbf {\bibinfo {volume} {49}},\ \bibinfo
  {pages} {405} (\bibinfo {year} {1982})}\BibitemShut {NoStop}%
\bibitem [{\citenamefont {Kohmoto}(1985)}]{Kohmoto1985}%
  \BibitemOpen
  \bibfield  {author} {\bibinfo {author} {\bibfnamefont {M.}~\bibnamefont
  {Kohmoto}},\ }\href {\doibase 10.1016/0003-4916(85)90148-4} {\bibfield
  {journal} {\bibinfo  {journal} {Annals of Physics}\ }\textbf {\bibinfo
  {volume} {160}},\ \bibinfo {pages} {343 } (\bibinfo {year}
  {1985})}\BibitemShut {NoStop}%
\bibitem [{\citenamefont {Kohmoto}(1989)}]{Kohmoto1989}%
  \BibitemOpen
  \bibfield  {author} {\bibinfo {author} {\bibfnamefont {M.}~\bibnamefont
  {Kohmoto}},\ }\href {\doibase 10.1103/PhysRevB.39.11943} {\bibfield
  {journal} {\bibinfo  {journal} {Phys. Rev. B}\ }\textbf {\bibinfo {volume}
  {39}},\ \bibinfo {pages} {11943} (\bibinfo {year} {1989})}\BibitemShut
  {NoStop}%
\bibitem [{\citenamefont {Lopes~dos Santos}\ \emph {et~al.}(2007)\citenamefont
  {Lopes~dos Santos}, \citenamefont {Peres},\ and\ \citenamefont
  {Castro~Neto}}]{Lopes2007}%
  \BibitemOpen
  \bibfield  {author} {\bibinfo {author} {\bibfnamefont {J.~M.~B.}\
  \bibnamefont {Lopes~dos Santos}}, \bibinfo {author} {\bibfnamefont
  {N.~M.~R.}\ \bibnamefont {Peres}}, \ and\ \bibinfo {author} {\bibfnamefont
  {A.~H.}\ \bibnamefont {Castro~Neto}},\ }\href {\doibase
  10.1103/PhysRevLett.99.256802} {\bibfield  {journal} {\bibinfo  {journal}
  {Phys. Rev. Lett.}\ }\textbf {\bibinfo {volume} {99}},\ \bibinfo {pages}
  {256802} (\bibinfo {year} {2007})}\BibitemShut {NoStop}%
\bibitem [{\citenamefont {Hass}\ \emph {et~al.}(2008)\citenamefont {Hass},
  \citenamefont {Varchon}, \citenamefont {Mill\'an-Otoya}, \citenamefont
  {Sprinkle}, \citenamefont {Sharma}, \citenamefont {de~Heer}, \citenamefont
  {Berger}, \citenamefont {First}, \citenamefont {Magaud},\ and\ \citenamefont
  {Conrad}}]{Hass2008}%
  \BibitemOpen
  \bibfield  {author} {\bibinfo {author} {\bibfnamefont {J.}~\bibnamefont
  {Hass}}, \bibinfo {author} {\bibfnamefont {F.}~\bibnamefont {Varchon}},
  \bibinfo {author} {\bibfnamefont {J.~E.}\ \bibnamefont {Mill\'an-Otoya}},
  \bibinfo {author} {\bibfnamefont {M.}~\bibnamefont {Sprinkle}}, \bibinfo
  {author} {\bibfnamefont {N.}~\bibnamefont {Sharma}}, \bibinfo {author}
  {\bibfnamefont {W.~A.}\ \bibnamefont {de~Heer}}, \bibinfo {author}
  {\bibfnamefont {C.}~\bibnamefont {Berger}}, \bibinfo {author} {\bibfnamefont
  {P.~N.}\ \bibnamefont {First}}, \bibinfo {author} {\bibfnamefont
  {L.}~\bibnamefont {Magaud}}, \ and\ \bibinfo {author} {\bibfnamefont {E.~H.}\
  \bibnamefont {Conrad}},\ }\href {\doibase 10.1103/PhysRevLett.100.125504}
  {\bibfield  {journal} {\bibinfo  {journal} {Phys. Rev. Lett.}\ }\textbf
  {\bibinfo {volume} {100}},\ \bibinfo {pages} {125504} (\bibinfo {year}
  {2008})}\BibitemShut {NoStop}%
\bibitem [{\citenamefont {Shallcross}\ \emph {et~al.}(2008)\citenamefont
  {Shallcross}, \citenamefont {Sharma},\ and\ \citenamefont
  {Pankratov}}]{Shallcross2008}%
  \BibitemOpen
  \bibfield  {author} {\bibinfo {author} {\bibfnamefont {S.}~\bibnamefont
  {Shallcross}}, \bibinfo {author} {\bibfnamefont {S.}~\bibnamefont {Sharma}},
  \ and\ \bibinfo {author} {\bibfnamefont {O.~A.}\ \bibnamefont {Pankratov}},\
  }\href {\doibase 10.1103/PhysRevLett.101.056803} {\bibfield  {journal}
  {\bibinfo  {journal} {Phys. Rev. Lett.}\ }\textbf {\bibinfo {volume} {101}},\
  \bibinfo {pages} {056803} (\bibinfo {year} {2008})}\BibitemShut {NoStop}%
\bibitem [{\citenamefont {Mele}(2010)}]{Mele2010}%
  \BibitemOpen
  \bibfield  {author} {\bibinfo {author} {\bibfnamefont {E.~J.}\ \bibnamefont
  {Mele}},\ }\href {\doibase 10.1103/PhysRevB.81.161405} {\bibfield  {journal}
  {\bibinfo  {journal} {Phys. Rev. B}\ }\textbf {\bibinfo {volume} {81}},\
  \bibinfo {pages} {161405} (\bibinfo {year} {2010})}\BibitemShut {NoStop}%
\bibitem [{\citenamefont {Trambly~de Laissardi\`ere}\ \emph
  {et~al.}(2010)\citenamefont {Trambly~de Laissardi\`ere}, \citenamefont
  {Mayou},\ and\ \citenamefont {Magaud}}]{Trambly2010}%
  \BibitemOpen
  \bibfield  {author} {\bibinfo {author} {\bibfnamefont {G.}~\bibnamefont
  {Trambly~de Laissardi\`ere}}, \bibinfo {author} {\bibfnamefont
  {D.}~\bibnamefont {Mayou}}, \ and\ \bibinfo {author} {\bibfnamefont
  {L.}~\bibnamefont {Magaud}},\ }\href {\doibase 10.1021/nl902948m} {\bibfield
  {journal} {\bibinfo  {journal} {Nano Letters}\ }\textbf {\bibinfo {volume}
  {10}},\ \bibinfo {pages} {804} (\bibinfo {year} {2010})}\BibitemShut
  {NoStop}%
\bibitem [{\citenamefont {Bistritzer}\ and\ \citenamefont
  {MacDonald}(2011{\natexlab{a}})}]{BistritzerPNAS2011}%
  \BibitemOpen
  \bibfield  {author} {\bibinfo {author} {\bibfnamefont {R.}~\bibnamefont
  {Bistritzer}}\ and\ \bibinfo {author} {\bibfnamefont {A.~H.}\ \bibnamefont
  {MacDonald}},\ }\href {\doibase 10.1073/pnas.1108174108} {\bibfield
  {journal} {\bibinfo  {journal} {Proceedings of the National Academy of
  Sciences}\ }\textbf {\bibinfo {volume} {108}},\ \bibinfo {pages} {12233}
  (\bibinfo {year} {2011}{\natexlab{a}})},\ \Eprint
  {http://arxiv.org/abs/http://www.pnas.org/content/108/30/12233.full.pdf+html}
  {http://www.pnas.org/content/108/30/12233.full.pdf+html} \BibitemShut
  {NoStop}%
\bibitem [{\citenamefont {Ponomarenko}\ \emph {et~al.}(2013)\citenamefont
  {Ponomarenko}, \citenamefont {Gorbachev}, \citenamefont {Yu}, \citenamefont
  {Elias}, \citenamefont {Jalil}, \citenamefont {Patel}, \citenamefont
  {Mishchenko}, \citenamefont {Mayorov}, \citenamefont {Woods}, \citenamefont
  {Wallbank}, \citenamefont {Mucha-Kruczynski}, \citenamefont {Piot},
  \citenamefont {Potemski}, \citenamefont {Grigorieva}, \citenamefont
  {Novoselov}, \citenamefont {Guinea}, \citenamefont {Fal'ko},\ and\
  \citenamefont {Geim}}]{Ponomarenko2012}%
  \BibitemOpen
  \bibfield  {author} {\bibinfo {author} {\bibfnamefont {L.~A.}\ \bibnamefont
  {Ponomarenko}}, \bibinfo {author} {\bibfnamefont {R.~V.}\ \bibnamefont
  {Gorbachev}}, \bibinfo {author} {\bibfnamefont {G.~L.}\ \bibnamefont {Yu}},
  \bibinfo {author} {\bibfnamefont {D.~C.}\ \bibnamefont {Elias}}, \bibinfo
  {author} {\bibfnamefont {R.}~\bibnamefont {Jalil}}, \bibinfo {author}
  {\bibfnamefont {A.~A.}\ \bibnamefont {Patel}}, \bibinfo {author}
  {\bibfnamefont {A.}~\bibnamefont {Mishchenko}}, \bibinfo {author}
  {\bibfnamefont {A.~S.}\ \bibnamefont {Mayorov}}, \bibinfo {author}
  {\bibfnamefont {C.~R.}\ \bibnamefont {Woods}}, \bibinfo {author}
  {\bibfnamefont {J.~R.}\ \bibnamefont {Wallbank}}, \bibinfo {author}
  {\bibfnamefont {M.}~\bibnamefont {Mucha-Kruczynski}}, \bibinfo {author}
  {\bibfnamefont {B.~A.}\ \bibnamefont {Piot}}, \bibinfo {author}
  {\bibfnamefont {M.}~\bibnamefont {Potemski}}, \bibinfo {author}
  {\bibfnamefont {I.~V.}\ \bibnamefont {Grigorieva}}, \bibinfo {author}
  {\bibfnamefont {K.~S.}\ \bibnamefont {Novoselov}}, \bibinfo {author}
  {\bibfnamefont {F.}~\bibnamefont {Guinea}}, \bibinfo {author} {\bibfnamefont
  {V.~I.}\ \bibnamefont {Fal'ko}}, \ and\ \bibinfo {author} {\bibfnamefont
  {A.~K.}\ \bibnamefont {Geim}},\ }\href@noop {} {\bibfield  {journal}
  {\bibinfo  {journal} {Nature (London)}\ }\textbf {\bibinfo {volume} {497}},\
  \bibinfo {pages} {594 } (\bibinfo {year} {2013})}\BibitemShut {NoStop}%
\bibitem [{\citenamefont {Dean}\ \emph {et~al.}(2013)\citenamefont {Dean},
  \citenamefont {Wang}, \citenamefont {Maher}, \citenamefont {Forsythe},
  \citenamefont {andY. Gao}, \citenamefont {Katoch}, \citenamefont {Ishigami},
  \citenamefont {Moon}, \citenamefont {Koshino}, \citenamefont {Taniguchi},
  \citenamefont {Watanabe}, \citenamefont {Shepard}, \citenamefont {Hone},\
  and\ \citenamefont {Kim}}]{Dean2012}%
  \BibitemOpen
  \bibfield  {author} {\bibinfo {author} {\bibfnamefont {C.~R.}\ \bibnamefont
  {Dean}}, \bibinfo {author} {\bibfnamefont {L.}~\bibnamefont {Wang}}, \bibinfo
  {author} {\bibfnamefont {P.}~\bibnamefont {Maher}}, \bibinfo {author}
  {\bibfnamefont {C.}~\bibnamefont {Forsythe}}, \bibinfo {author}
  {\bibfnamefont {F.~G.}\ \bibnamefont {andY. Gao}}, \bibinfo {author}
  {\bibfnamefont {J.}~\bibnamefont {Katoch}}, \bibinfo {author} {\bibfnamefont
  {M.}~\bibnamefont {Ishigami}}, \bibinfo {author} {\bibfnamefont
  {P.}~\bibnamefont {Moon}}, \bibinfo {author} {\bibfnamefont {M.}~\bibnamefont
  {Koshino}}, \bibinfo {author} {\bibfnamefont {T.}~\bibnamefont {Taniguchi}},
  \bibinfo {author} {\bibfnamefont {K.}~\bibnamefont {Watanabe}}, \bibinfo
  {author} {\bibfnamefont {K.~L.}\ \bibnamefont {Shepard}}, \bibinfo {author}
  {\bibfnamefont {J.}~\bibnamefont {Hone}}, \ and\ \bibinfo {author}
  {\bibfnamefont {P.}~\bibnamefont {Kim}},\ }\href@noop {} {\bibfield
  {journal} {\bibinfo  {journal} {Nature (London)}\ }\textbf {\bibinfo {volume}
  {497}},\ \bibinfo {pages} {598 } (\bibinfo {year} {2013})}\BibitemShut
  {NoStop}%
\bibitem [{\citenamefont {Lee}\ \emph {et~al.}(2011)\citenamefont {Lee},
  \citenamefont {Riedl}, \citenamefont {Beringer}, \citenamefont {Castro~Neto},
  \citenamefont {von Klitzing}, \citenamefont {Starke},\ and\ \citenamefont
  {Smet}}]{Lee2011}%
  \BibitemOpen
  \bibfield  {author} {\bibinfo {author} {\bibfnamefont {D.~S.}\ \bibnamefont
  {Lee}}, \bibinfo {author} {\bibfnamefont {C.}~\bibnamefont {Riedl}}, \bibinfo
  {author} {\bibfnamefont {T.}~\bibnamefont {Beringer}}, \bibinfo {author}
  {\bibfnamefont {A.~H.}\ \bibnamefont {Castro~Neto}}, \bibinfo {author}
  {\bibfnamefont {K.}~\bibnamefont {von Klitzing}}, \bibinfo {author}
  {\bibfnamefont {U.}~\bibnamefont {Starke}}, \ and\ \bibinfo {author}
  {\bibfnamefont {J.~H.}\ \bibnamefont {Smet}},\ }\href {\doibase
  10.1103/PhysRevLett.107.216602} {\bibfield  {journal} {\bibinfo  {journal}
  {Phys. Rev. Lett.}\ }\textbf {\bibinfo {volume} {107}},\ \bibinfo {pages}
  {216602} (\bibinfo {year} {2011})}\BibitemShut {NoStop}%
\bibitem [{\citenamefont {Bistritzer}\ and\ \citenamefont
  {MacDonald}(2011{\natexlab{b}})}]{BistritzerPRB2011}%
  \BibitemOpen
  \bibfield  {author} {\bibinfo {author} {\bibfnamefont {R.}~\bibnamefont
  {Bistritzer}}\ and\ \bibinfo {author} {\bibfnamefont {A.~H.}\ \bibnamefont
  {MacDonald}},\ }\href {\doibase 10.1103/PhysRevB.84.035440} {\bibfield
  {journal} {\bibinfo  {journal} {Phys. Rev. B}\ }\textbf {\bibinfo {volume}
  {84}},\ \bibinfo {pages} {035440} (\bibinfo {year}
  {2011}{\natexlab{b}})}\BibitemShut {NoStop}%
\bibitem [{\citenamefont {Moon}\ and\ \citenamefont
  {Koshino}(2012)}]{Moon2012}%
  \BibitemOpen
  \bibfield  {author} {\bibinfo {author} {\bibfnamefont {P.}~\bibnamefont
  {Moon}}\ and\ \bibinfo {author} {\bibfnamefont {M.}~\bibnamefont {Koshino}},\
  }\href {\doibase 10.1103/PhysRevB.85.195458} {\bibfield  {journal} {\bibinfo
  {journal} {Phys. Rev. B}\ }\textbf {\bibinfo {volume} {85}},\ \bibinfo
  {pages} {195458} (\bibinfo {year} {2012})}\BibitemShut {NoStop}%
\bibitem [{\citenamefont {Wang}\ \emph {et~al.}(2012)\citenamefont {Wang},
  \citenamefont {Liu},\ and\ \citenamefont {Chou}}]{Wang2012}%
  \BibitemOpen
  \bibfield  {author} {\bibinfo {author} {\bibfnamefont {Z.~F.}\ \bibnamefont
  {Wang}}, \bibinfo {author} {\bibfnamefont {F.}~\bibnamefont {Liu}}, \ and\
  \bibinfo {author} {\bibfnamefont {M.~Y.}\ \bibnamefont {Chou}},\ }\href
  {\doibase 10.1021/nl301794t} {\bibfield  {journal} {\bibinfo  {journal} {Nano
  Letters}\ }\textbf {\bibinfo {volume} {12}},\ \bibinfo {pages} {3833}
  (\bibinfo {year} {2012})}\BibitemShut {NoStop}%
\bibitem [{\citenamefont {Poilblanc}\ \emph {et~al.}(1990)\citenamefont
  {Poilblanc}, \citenamefont {Hasegawa},\ and\ \citenamefont
  {Rice}}]{Poilblanc1990}%
  \BibitemOpen
  \bibfield  {author} {\bibinfo {author} {\bibfnamefont {D.}~\bibnamefont
  {Poilblanc}}, \bibinfo {author} {\bibfnamefont {Y.}~\bibnamefont {Hasegawa}},
  \ and\ \bibinfo {author} {\bibfnamefont {T.~M.}\ \bibnamefont {Rice}},\
  }\href {\doibase 10.1103/PhysRevB.41.1949} {\bibfield  {journal} {\bibinfo
  {journal} {Phys. Rev. B}\ }\textbf {\bibinfo {volume} {41}},\ \bibinfo
  {pages} {1949} (\bibinfo {year} {1990})}\BibitemShut {NoStop}%
\bibitem [{\citenamefont {Hatsugai}\ \emph {et~al.}(1999)\citenamefont
  {Hatsugai}, \citenamefont {Ishibashi},\ and\ \citenamefont
  {Morita}}]{Hatsugai1999}%
  \BibitemOpen
  \bibfield  {author} {\bibinfo {author} {\bibfnamefont {Y.}~\bibnamefont
  {Hatsugai}}, \bibinfo {author} {\bibfnamefont {K.}~\bibnamefont {Ishibashi}},
  \ and\ \bibinfo {author} {\bibfnamefont {Y.}~\bibnamefont {Morita}},\ }\href
  {\doibase 10.1103/PhysRevLett.83.2246} {\bibfield  {journal} {\bibinfo
  {journal} {Phys. Rev. Lett.}\ }\textbf {\bibinfo {volume} {83}},\ \bibinfo
  {pages} {2246} (\bibinfo {year} {1999})}\BibitemShut {NoStop}%
\bibitem [{\citenamefont {Trellakis}(2003)}]{Trellakis2003}%
  \BibitemOpen
  \bibfield  {author} {\bibinfo {author} {\bibfnamefont {A.}~\bibnamefont
  {Trellakis}},\ }\href {\doibase 10.1103/PhysRevLett.91.056405} {\bibfield
  {journal} {\bibinfo  {journal} {Phys. Rev. Lett.}\ }\textbf {\bibinfo
  {volume} {91}},\ \bibinfo {pages} {056405} (\bibinfo {year}
  {2003})}\BibitemShut {NoStop}%
\bibitem [{\citenamefont {Cai}\ and\ \citenamefont {Galli}(2004)}]{Cai2004}%
  \BibitemOpen
  \bibfield  {author} {\bibinfo {author} {\bibfnamefont {W.}~\bibnamefont
  {Cai}}\ and\ \bibinfo {author} {\bibfnamefont {G.}~\bibnamefont {Galli}},\
  }\href {\doibase 10.1103/PhysRevLett.92.186402} {\bibfield  {journal}
  {\bibinfo  {journal} {Phys. Rev. Lett.}\ }\textbf {\bibinfo {volume} {92}},\
  \bibinfo {pages} {186402} (\bibinfo {year} {2004})}\BibitemShut {NoStop}%
\bibitem [{\citenamefont {Nakanishi}\ and\ \citenamefont
  {Ando}(2001)}]{Nakanishi2001}%
  \BibitemOpen
  \bibfield  {author} {\bibinfo {author} {\bibfnamefont {T.}~\bibnamefont
  {Nakanishi}}\ and\ \bibinfo {author} {\bibfnamefont {T.}~\bibnamefont
  {Ando}},\ }\href {\doibase 10.1143/JPSJ.70.1647} {\bibfield  {journal}
  {\bibinfo  {journal} {Journal of the Physical Society of Japan}\ }\textbf
  {\bibinfo {volume} {70}},\ \bibinfo {pages} {1647} (\bibinfo {year}
  {2001})}\BibitemShut {NoStop}%
\bibitem [{\citenamefont {Rhim}\ and\ \citenamefont {Park}(2012)}]{Rhim2012}%
  \BibitemOpen
  \bibfield  {author} {\bibinfo {author} {\bibfnamefont {J.-W.}\ \bibnamefont
  {Rhim}}\ and\ \bibinfo {author} {\bibfnamefont {K.}~\bibnamefont {Park}},\
  }\href {\doibase 10.1103/PhysRevB.86.235411} {\bibfield  {journal} {\bibinfo
  {journal} {Phys. Rev. B}\ }\textbf {\bibinfo {volume} {86}},\ \bibinfo
  {pages} {235411} (\bibinfo {year} {2012})}\BibitemShut {NoStop}%
\bibitem [{\citenamefont {Sato}\ \emph {et~al.}(2008)\citenamefont {Sato},
  \citenamefont {Tobe},\ and\ \citenamefont {Kohmoto}}]{Sato2008}%
  \BibitemOpen
  \bibfield  {author} {\bibinfo {author} {\bibfnamefont {M.}~\bibnamefont
  {Sato}}, \bibinfo {author} {\bibfnamefont {D.}~\bibnamefont {Tobe}}, \ and\
  \bibinfo {author} {\bibfnamefont {M.}~\bibnamefont {Kohmoto}},\ }\href
  {\doibase 10.1103/PhysRevB.78.235322} {\bibfield  {journal} {\bibinfo
  {journal} {Phys. Rev. B}\ }\textbf {\bibinfo {volume} {78}},\ \bibinfo
  {pages} {235322} (\bibinfo {year} {2008})}\BibitemShut {NoStop}%
\end{thebibliography}%

\end{document}